\documentclass[%
 reprint,
 amsmath,amssymb,
 aps,
]{revtex4-2}
\usepackage{lipsum}
\usepackage{graphicx}
\usepackage{dcolumn}
\usepackage{bm}
\usepackage{color}
\usepackage{comment}
\usepackage{physics}
\usepackage{blindtext}

\begin{document}
\preprint{APS/123-QED}
\title{ 3/2 magic-angle quantization rule of flat bands in
twisted bilayer graphene and relationship with the Quantum Hall effect}
\author{Leonardo A. Navarro-Labastida and Gerardo G. Naumis}
\date{February 2023}
\email{naumis@fisica.unam.mx}
\affiliation{%
Depto. de Sistemas Complejos, Instituto de F\'isica, \\ Universidad Nacional Aut\'onoma de M\'exico (UNAM)\\
Apdo. Postal 20-364, 01000, CDMX, M\'exico.
}%
\begin{abstract}
Flat band electronic modes in twisted graphene bilayers are responsible for superconducting and other highly  correlated electron-electron phases. Although some hints were known of a possible connection between the quantum Hall effect and zero flat band modes, it was not clear how such connection appears. Here the electronic behavior in twisted bilayer graphene is studied using the chiral model Hamiltonian. As a result, it is proved that for high-order magic angles, the zero flat band modes converge into coherent Landau states with a dispersion  $\sigma^2=1/3\alpha$, where $\alpha$ is a coupling parameter that incorporates the twist angle and energetic scales.  Then it is proved that the square of the hamiltonian, which is a $2\times 2$ matrix operator, turns out to be equivalent in a first approximation to a two-dimensional quantum harmonic oscillator. The interlayer currents between graphene's bipartite lattices are identified with the angular momentum term while the confinement potential is an effective quadratic potential. By considering the zero mode equation, the boundary conditions and a scaling argument, a limiting quantization rule for high-order magic angles is obtained, i.e., $\alpha_{m+1}-\alpha_{m}=3/2$ where $m$ is the order of the angle. From there, an equipartition and quantization of the kinetic, confinement and angular momentum contributions is found. All these results are in very good agreement with numerical calculations.  
\end{abstract}

\maketitle
\section{Introduction}
In 2018 it was found experimentally that twisted bilayer graphene (TBG) presents strongly correlated electron-electron quantum phases leading for example to unconventional superconductivity and Mott insulator states \cite{Cao2018}. More recently, trilayer twisted graphene has been found to be the most strongly interacting correlated material \cite{park2021,2022Ashvin}. Such remarkable discoveries presented a new paradigm in the so-called Moiré materials and unveiled the importance of two-dimensional (2D) materials to understand unconventional superconductivity in cuprates and heavy fermions systems, as they share similar quantum phase diagrams \cite{Cao2018, park2021,Bernevig2022}. TBG  advantages are i) its simplicity, as they are made from a single chemical element, and ii) they have a high degree of manipulation that cuprates doesn't have.  In recent years, there has been a significant interest in these phases of matter from a fundamental point of view \cite{Zachary2019_2,Hridis2019,Kerelsky2019,Fu2020,ochoa2021degradation,Namm2020,Oka2021,Manato2021,Rubio2021}  but also because they present a  lot of possible electronic applications and quantum computing advantages \cite{2021Rochee,Rubio2021}.  There is also an interesting connection between topological phases, edge states, semimetals, and fractional quantum Hall effect (FQHE) \cite{Barth2020,2021ShangL,2021Liu,2022Ledwith_Vortex,2022Vishwanath,Xu2018,2019FengCheng,Stauber2020,2018Wu,2018Low,2022Pantaleon,2020Yixing}. A recently paper establishes a connection between heavy fermion models and TBG \cite{Bernevig2022}, opening the prospect of using heavy fermions physics to the superconducting physics of TBG and more strongly correlated phases. \\

The discovery of such phases was proceeded by the Bistritzer-Mac Donald (BM) theoretical observation that twisted bilayer graphene (TBG) develops flat bands at certain twisting angles which are called magic \cite{MacDonald2011}. BM considered a continuum Dirac model in which the moiré periodicity between layers produces moiré Bloch´s bands  \cite{MacDonald2011}. The model is continuum in the sense that the interlayer potential between Carbon $\pi$  orbitals is a smooth function of the spatial separation projected onto the graphene planes and also the hopping is local and periodic, allowing to apply the Bloch´s theorem for any rotation angle. For TBG it was demonstrated that non-Abelian gauge fields arise due to the coupling between layers in the low-energy regime \cite{Guinea2012,Naumis2021}. 

Flat band modes that arise at magic angles, also known as zero energy modes, have been investigated in many recent works \cite{Tarnpolsky2019,Ledwidth2020,Ledwith2021, WANGG2021,CANO2021,popovF2021,2018ChengCheng,2022Onari,Saul2021,Patrick2021}, and in particular, there were hints in the mathematics for a possible connection with the quantum Hall effect (QHE) and the lowest Landau level \cite{Tarnpolsky2019,WANGG2021}.  There are interesting properties of the zero mode wave function \cite{WANGG2021,CANO2021,popovF2021}, in particular, the connection with the lowest Landau level reveals that TBG presents topological phases \cite{WANGG2021,2021yardenn}.  
 
Importantly, the wave function is reminiscent of a quantum hall wave function because is described in terms of Jacobi theta functions such as in the quantum hall effect wave function \cite{Tarnpolsky2019,Ledwidth2020,Ledwith2021}.  This hidden wave function is important to understand because leads to particular localization properties, orbital current, density wave function distribution, and symmetries of the pseudo-magnetic gauge fields. Yet, exactly how this analogy arises was not clear as no  connection between the quantum harmonic oscillator and the TBG hamiltonian  was ever found.  Tarnopolsky et. al. also found that magic angles were quantized but no explanation was provided for this fact \cite{Tarnpolsky2019}. Thus there were
two open questions related to the same problem. The present work shows how these two questions relate to each other, and also answers them. Moreover, we find that in fact, the zero flat band modes converge into coherent Landau levels. As we will discuss, this is done by using boundary layer differential equations theory and squaring the Hamiltonian \cite{Naumis2021,Naumis2022}, a process that has also been used in supersymmetry \cite{2022Mizoguchi,Hatsugai2021,2022Krishanu}.\\

\section{Chiral TBG and Squared TBG Hamiltonians}
The chiral Hamiltonian of twisted bilayer graphene is a variant of the original Bistritzer-MacDonald Hamiltonian in which the $AA$ tunneling is set to zero \cite{Ledwith2021}. Here we use as basis the wave vectors  $\Phi(\bm{r})=\begin{pmatrix} 
\psi_1(\bm{r}) ,
\psi_2(\bm{r}),
\chi_1(\bm{r}),
\chi_2(\bm{r})
\end{pmatrix}^T$ where the index $1,2$ represents each graphene layer and $\psi_j(\bm{r})$ and $\chi_j(\bm{r})$ are the Wannier orbitals on each inequivalent site of the graphene's unit cell. 
The chiral Hamiltonian is given by \cite{Tarnpolsky2019,Khalaff2019, Ledwidth2020}, 
\begin{equation}
\begin{split}
\mathcal{H}
&=\begin{pmatrix} 
0 & D^{\ast}(-\bm{r})\\
 D(\bm{r}) & 0
  \end{pmatrix}  \\
\end{split} 
\label{H_initial}
\end{equation}
where the zero-mode operator is defined as, 
\begin{equation}
\begin{split}
D(\bm{r})&=\begin{pmatrix} 
-i\Bar{\partial} & \alpha U(\bm{r})\\
  \alpha U(-\bm{r}) & -i\Bar{\partial} 
  \end{pmatrix}  \\
\end{split} 
\end{equation}
and,
\begin{equation}
\begin{split}
D^{*}(-\bm{r})&=\begin{pmatrix} 
-i\partial & \alpha U^{*}(-\bm{r})\\
  \alpha U^{*}(\bm{r}) & -i\partial 
  \end{pmatrix}  \\
\end{split} 
\end{equation}
with $\Bar{\partial}=\partial_x+i\partial_y$, $\partial=\partial_x-i\partial_y$. The potential is,
{\footnotesize
\begin{equation}
    U(\bm{\bm{r}})=e^{-i\bm{q}_1\cdot \bm{r}}+e^{i\phi}e^{-i\bm{q}_2\cdot \bm{r}}+e^{-i\phi}e^{-i\bm{q}_3\cdot \boldsymbol{r}}
\end{equation}
}
where the phase factor is $\phi=2\pi/3$ and the vectors are given by $\bm{q}_{1}=k_{\theta}(0,-1)$, $\bm{q}_{2}=k_{\theta}(\frac{\sqrt{3}}{2},\frac{1}{2})$, $\bm{q}_{3}=k_{\theta}(-\frac{\sqrt{3}}{2},\frac{1}{2})$, the moir\'e modulation vector is $k_{\theta}=2k_{D}\sin{\frac{\theta}{2}}$ with
$k_{D}=\frac{4\pi}{3a_{0}}$ is the magnitude of the Dirac wave vector
and $a_{0}$ is the lattice constant of monolayer graphene. The model contains only the parameter $\alpha$, defined as $\alpha=\frac{w_1}{v_0 k_\theta}$ where $w_1$ is the interlayer coupling of stacking AB/BA with value $w_1=110$ meV and $v_0$ is the Fermi velocity with value $v_0=\frac{19.81eV}{2k_D}$.
The operators $\partial$ and $\Bar{\partial}$ are dimensionless as the Hamiltonian Eq. (\ref{H_initial}) is written in using units where $v_0=1$, $k_{\theta}=1$. The twist angle only enters in the dimensionless parameter $\alpha$.  The combinations $\bm{b}_{1,2}=\bm{q}_{2,3}-\bm{q}_{1}$ are the moir\'e Brillouin zone (mBZ)  vectors and also $\bm{b}_{3}=\bm{q}_{3}-\bm{q}_2$.
Using this basis for the reciprocal space lattice, some important high symmetry points of the moir\'e Brillouin zone are $\bm{K}=(0,0)$, $\bm{K'}=-\bm{q}_1$, and $\bm{\Gamma}=\bm{q}_1$ (see ref. \cite{Naumis2022} for a diagram). 
For further use it is also convenient to define  a set of unitary vectors $\bm{q}_{\mu}^{\perp}$ perpendicular to the set $\bm{q}_{\mu}$ and given by
$\bm{q}_{1}^{\perp}=(1,0),  \bm{q}_{2}^{\perp}=\big(-\frac{1}{2},\frac{\sqrt{3}}{2}\big), 
    \bm{q}_{3}^{\perp}=\big(-\frac{1}{2},-\frac{ \sqrt{3}}{2}\big)$.
The moir\'e vectors  unitary cell are given by $\bm{a}_{1,2}=(4\pi/3k_{\theta})(\sqrt{3}/2,1/2)$. Observe that $\bm{q}_{\mu}\cdot \bm{a}_{1,2}=-\phi$ for $\mu=1,2,3$.  
    

In a previous work  we showed how, by taking the square of $H$, it is possible to write the Hamiltonian as a $2 \times 2$ matrix \cite{Naumis2021,Naumis2022},  %
\begin{equation}\label{eq:H2}
\begin{split}
H^{2}&=\begin{pmatrix} -\nabla^{2}+\alpha^{2}|U(-\bm{r})|^{2} &  \alpha A^{\dagger}(\bm{r}) \\
 \alpha A(\bm{r}) & -\nabla^{2}+\alpha^{2}|U(\bm{r})|^{2} 
  \end{pmatrix} 
  \end{split} 
  \end{equation}
where the squared norm of the potential is an effective trigonal confinement potential,
\begin{equation}
\begin{split}
|U(\bm{r})|^{2} &= 3+2 \cos(\bm{b}_1\cdot \bm{r}-\phi)+2\cos(\bm{b}_2\cdot \bm{r}+\phi )\\
 & +2\cos(\bm{b}_3\cdot \bm{r}+2\phi)   
\end{split}
\end{equation}
and the off-diagonal term is,
\begin{equation}
 \begin{split}
A^{\dagger}(\bm{r}) & =-i\sum_{\mu=1}^{3}e^{-i\bm{q}_{\mu}\cdot\bm{r}}(2\bm{q}_{\mu}^{\perp}\cdot \bm{\nabla}+1)\\
  \end{split} 
\end{equation}
where  $\bm{\nabla}^{\dagger}=-\bm{\nabla}$ with $\bm{\nabla}=(\partial_x,\partial_y)$ and $\mu=1,2,3$.

\section{Zero-energy modes as coherent Landau states}

Now we investigate the asymptotic limit $\alpha\rightarrow\infty$ by numerically solving  (see appendix C) the Schrödinger equation $\mathcal{H}\Psi(\bm{r})=E\Psi(\bm{r})$ where $E$ is the energy. As the potential is periodic, it satisfies  Bolch's theorem, and thus $\psi_{\bm{k},j}(\bm{r})=e^{i\bm{k}\cdot \bm{r}}u_{\bm{k},j}(\bm{r})$ where $u_{\bm{k},j}(\bm{r})$ has the periodicity of the lattice (see appendix A). The rotational $C_3$ symmetry allows to further simplify the problem (see appendix B). In Fig. \ref{fig:Structure_Nodalpoints} we present
 the zero mode wave function, corresponding to $E=0$ at the reciprocal space  point $\bm{k}=\Gamma$  for the $m$th magic angles ($\alpha_m$) with $m=8$ and $m=9$.  The electronic maxima of the density form hexagons which are nearly localized at $\bm{r_{\mu}} \approx \pm \bm{q}_{\mu}$. Such observation is detailed in Fig. \ref{fig:Structure_Nodalpoints}. Moreover, the wave-function for other $ \bm{k}$ points follow the same behavior although the $\Gamma$ point best captures the magic angle behavior \cite{Naumis2022}. In the limit of $\alpha_{m} \rightarrow \infty$ we have verified that in fact, the electron density is almost localized at $\bm{q}_{\mu}$. Notice that here we are working with adimensional units but this suggests a connection with the QHE as solutions seem self-dual  \cite{Hofstadter_1976}, i.e., in real space are similar to those in reciprocal space  with renormalized parameters.

\begin{figure}[h!]
\includegraphics[scale=0.4]{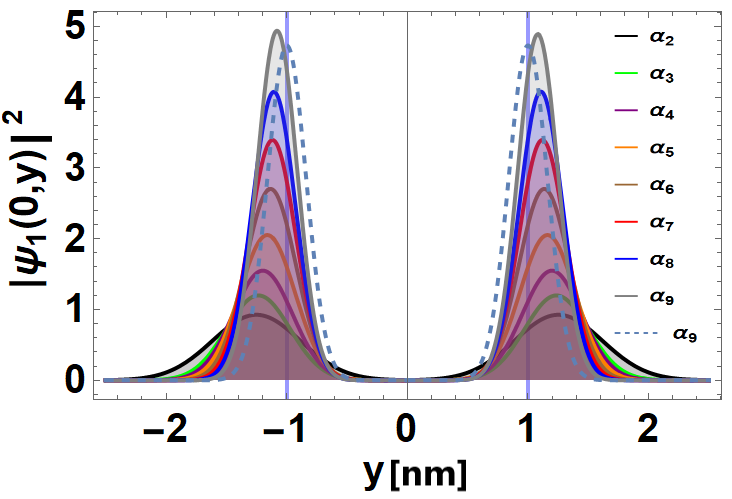}\caption{Coherent states limit for the wave function density along the $y$-axis for layer $1$.  The solid curves with shaded areas are the  normalized numerical solutions of the zero mode Eqns. (\ref{eq:eq1zeroA}) and (\ref{eq:eq2zeroB}) for the indicated magic angles. The dashed line is the normalized theoretical result using a coherent state using the ninth-magic angle ($\alpha_9$). The thick vertical lines (blue lines) indicate the limiting localization points $\bm{q}_1$ and $-\bm{q}_1$. Notice how as the magic angle goes from the second to the nine, the density becomes sharply peaked.}
\label{fig:GaussianLimit}
\end{figure}

\begin{figure}[h!]
\includegraphics[scale=0.9]{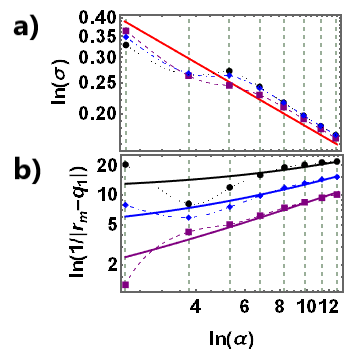}\caption{a) Log-log plot showing the standard deviation $\sigma$ of the zero energy modes (coherent Landau states) vs $\alpha_m$. The red solid line is the theoretical equation $\sigma=1/\sqrt{3\alpha}$ while the markers are obtained from a fit with Gaussians to the electronic density  obtained numerically from the zero mode equation at the $\bm{\Gamma}$ (purple), $\bm{K}$ (black) and $\bm{M}$ (blue) points of the Brillouin zone.  The short dashed lines joining markers are  used as a guide to the eye. An excellent agreement between the theory and the numerical results is seen for high order magic angles. b) Log-log plot of the inverse distance from $\bm{q}_1$ to $\bm{r}_m$, which is the position of the closest maximum of the numerically obtained electronic density.The solid curves visually show the asymptotic convergence $\bm{r}_m \rightarrow \bm{q}_1$ for all curves when $\alpha\rightarrow \infty$. For both panels, the green vertical dashed lines indicate magic angles from the second to the nine, showing the $\alpha_{m+1}-\alpha_{m} \approx 3/2$ rule for the magic angle separation.} \label{fig:Sigmaalpha}
\end{figure}

Although there are expressions for the wave-function \cite{Tarnpolsky2019,Wang2020,CANO2021} at any $\bm{k}$ point that hinted a relationship with the lowest Landau levels, they depend on the wave function at the $\bm{K}$ point, i.e., 
\begin{equation}
    \psi_{\bm{k},j}(\bm{r})=f_{\bm{k}}(z) \psi_{\bm{K},j}(\bm{r})
    \label{eq:fTarnopolsky}
\end{equation}
where $z=x+iy$ and $f_{\bm{k}}(z)$ is an analytic function which satisfy the boundary condition and turns out to be a Jacobi theta function. The form of the $\psi_{\bm{K},j}(\bm{r})$ is not analytically known.  Yet in Figs. \ref{fig:Structure_Nodalpoints} and \ref{fig:GaussianLimit} we see numerically that the electron wave function reaches an asymptotic limit almost invariant as $\alpha_m \rightarrow \infty$. In this limit, the localization centers for the $\bf{\Gamma}$ point wave function seem to converge as seen in Figs. \ref{fig:Structure_Nodalpoints}, \ref{fig:GaussianLimit} and \ref{fig:Sigmaalpha}. Such wave function tends to be localized in certain points of space which are not the stacking points AA, AB, and BA. In that sense, the solutions are very different from the first magic angle a fact that was explained elsewhere \cite{Naumis2022}. As seen in Fig. \ref{fig:Sigmaalpha}, for other $\bm{k}$ points different from $\bf{\Gamma}$ the situation is quite similar, i.e., the wave functions are more localized as $\alpha \rightarrow \infty$ and approach the same localization center. 

\begin{figure}[h!]
\includegraphics[scale=0.25]{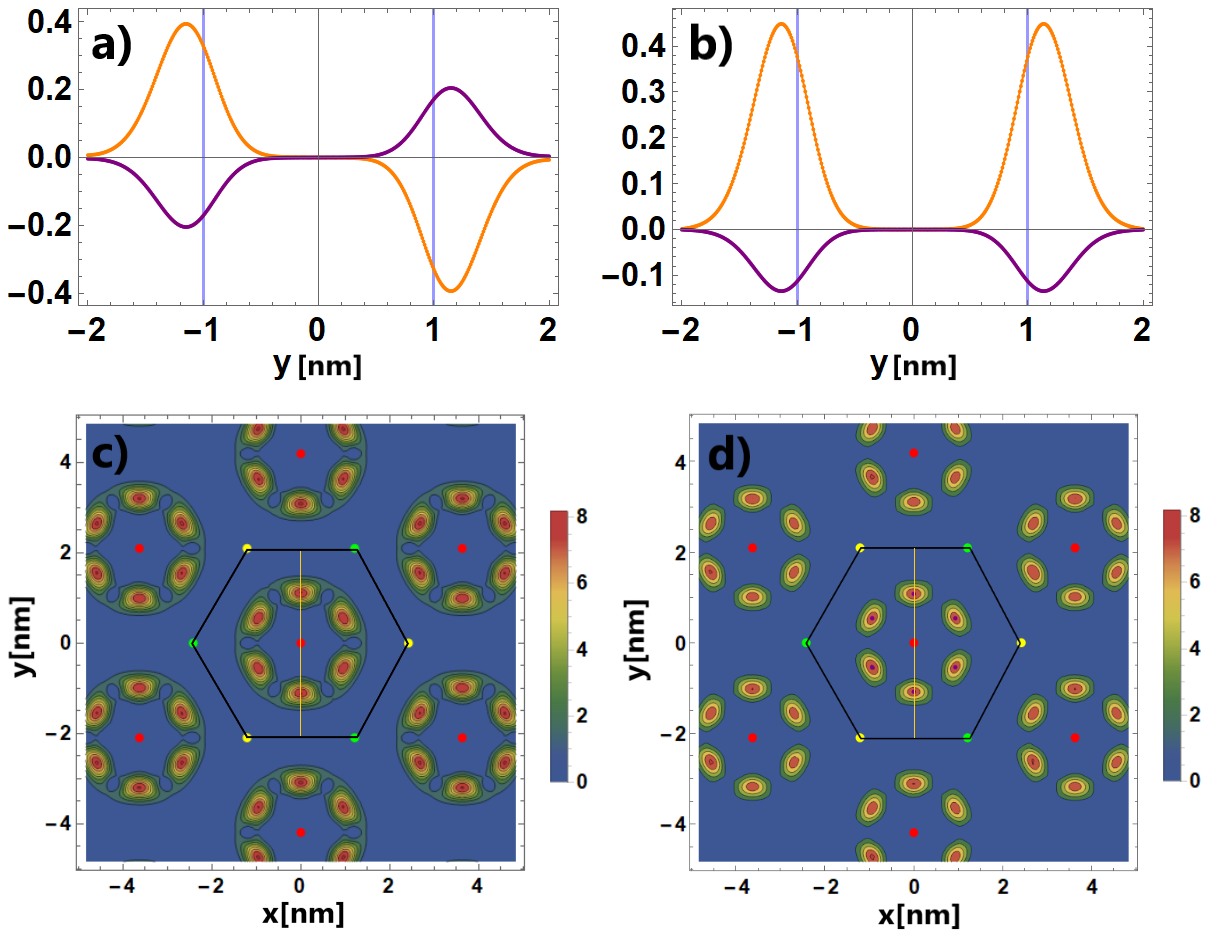}\caption{ Numerically obtained zero-mode normalized wave function localization for some high-order magic angles at $\bm{k}=\bm{\Gamma}$. In Panels a) and b) we present $\Re{\psi_1(\bm{r})}+\Im{\psi_2(\bm{r})}$ (orange curves) and $\Im{\psi_1(\bm{r})}-\Re{\psi_2(\bm{r})}$ (purple curves) parts of a one layer symmetrized wave function components $\psi_{\pm}(\bm{r})=\psi_1(\bm{r})\mp i\mu_{\alpha}\psi_2(\bm{r})$ at the symmetric line $(0,y)$ at magic angles $\alpha_8=11.345$ and $\alpha_9=12.855$ respectively.  Panels c) and d), contour plot of the global electronic density $\rho_1(\bm{r})+\rho_2(\bm{r})$ for $\alpha_{8}$ and  $\alpha_{9}$ respectively. The vertical line (yellow line) inside the real-space moire unit cell indicates the cut along the $y$ axis used in panels a) and b). The external hexagon is the real-space moire unit cell, where the AB (green), BA (yellow), and AA (red) stacking points are indicated. For higher magic angles, the wave-function density localizes in $6$ high-density points,  located at $\bm{r}=\pm\bm{q}_{\mu}$, with $\mu=1,2,3$, forming the red spots of maximal density.}
\label{fig:Structure_Nodalpoints}
\end{figure}

To understand how this limiting wave function arises, let us discuss the zero-mode equation $D(\bm{r})\begin{pmatrix} 
\psi_1(\bm{r}) ,
\psi_2(\bm{r})
\end{pmatrix}^T=0$ for states in the flat-band.
Although not essential for the analysis, it is easier to understand the $\Gamma$ point solution. For this case we have
that due to symmetry, $\psi_2(\bm{r})=i\mu_{\alpha} \psi_1(-\bm{r})$ where $\mu_{\alpha}=\pm 1$ depending on the magic angle parity \cite{Tarnpolsky2019}. 
Therefore, we obtain,
\begin{equation}\label{eq:eq1zeroA}
    \bar{\partial} \psi_1(\bm{r})= \alpha\mu_{\alpha}U(\bm{r}) \psi_1(-\bm{r})
\end{equation}
\begin{equation}\label{eq:eq2zeroB}
     \bar{\partial} \psi_1(-\bm{r})= -\alpha\mu_{\alpha} U(-\bm{r}) \psi_1(\bm{r})
\end{equation}
To solve the equation in the limit $\alpha \rightarrow \infty$
we use  the boundary layer theory of differential equations \cite{BookDE}, i.e., whenever the gradients are small, we can neglect the derivative in Eqns. (\ref{eq:eq1zeroA})-(\ref{eq:eq2zeroB}) when compared to the potential term. Then our solution must satisfy $\psi_j(r) \rightarrow 0$. The solution will be different from zero only inside the boundary layer, i.e., whenever $\bar \partial \psi_j(\bm{r})$ is of order $\alpha U(\pm \bm{r})\psi_j(\bm{r})$. Taken into account the boundary layer we conclude that the solution must be strongly peaked around certain regions of space. Then is natural to seek the solution within continuous functions having a peak while keeping the form of Eq. (\ref{eq:fTarnopolsky}).    
We then propose a coherent Landau state ansatz for a given layer (and thus suppress the subindex $j$),
\begin{equation}
\begin{split}\label{eq:coherent_psi}
\psi(z,z^*)&=f_{\lambda}(z)e^{-\frac{1}{4\sigma^2}|z|^2}\\
  \end{split}
  \end{equation}
where $f_\lambda(z)$ is an analytic function \cite{BookQM},
\begin{equation}
\begin{split}\label{eq:coherent_fz1}
f_{\lambda}(z)&=\frac{1}{\sigma\sqrt{2\pi}}e^{\frac{1}{2\sigma^2}\lambda^* z}e^{-\frac{1}{4\sigma^2}\lambda\lambda^*}
  \end{split} 
  \end{equation}
The parameter $\lambda$ is the localization center (known as the guiding coordinates in the QHE problem \cite{BookGrivin}) and $\sigma$ the standard deviation as the  electronic density is a Gaussian,
\begin{equation}
    \rho(\bm{r})=\frac{1}{2\pi\sigma^2}e^{\frac{-|z-\lambda|^2}{2\sigma^2}}
\end{equation}
Notice how the Gaussian envelope in Eq. (\ref{eq:coherent_psi}) ensures the boundary layer condition, i.e., the vanishing of the wave function whenever the gradient is small.

However, still we need to make remarks. As the equation involves $\psi(\bm{r})$ and $\psi(-\bm{r})$, the solutions can be written as a sum of a symmetrized and antisymmetrized forms. Therefore, it will be a linear combination of the symmetrized/antisymmetrized wavefunctions, 
\begin{equation}
\begin{split}\label{eq:psi_coherentState1}
\psi_{\pm}(z,z^*) \approx e^{-\frac{1}{4\sigma^2}|z|^2}\frac{1}{\sqrt{2}}(f_{\lambda}(z)\pm  f_{-\lambda}(z))\\
 \end{split}
 \end{equation}
provided that $\sigma \rightarrow \infty$ to avoid overlap between the Gaussians centered at $\lambda$ and $-\lambda$. A second reason to neglect the overlap effect around $z=0$ is that $U(0)=0$ and $\overline \partial \psi_{\pm}(z,z^*)|_{z,z^*=0}=0$. 

In what follows we will use our ansatz in the zero mode equation to prove how it satisfies the equation and to obtain $\sigma$. 

Before doing so, observe that $\psi(r)$ must transform according to the $C_3$ symmetry group and this can be ensured by defining a $\lambda_1$ such that, 


\begin{equation}
\begin{split}\label{eq:psi_coherentState2}
\psi_{\pm}(z,z^*) \approx \frac{1}{{\sqrt{6}}}\sum_{\mu=1}^3e^{-\frac{1}{4\sigma^2}|z|^2}(f_{\lambda_{\mu}}(z)\pm  f_{-\lambda_{\mu}}(z))\\
 \end{split}
\end{equation}
 where $\lambda_{2}=e^{i\phi}\lambda_{1}$ and  $\lambda_{3}=e^{-i\phi}\lambda_{1}$ and the normalization constant was modified to account for two layers and the three $\lambda_{\mu}$. Concerning the boundary conditions, i.e., the Bloch's theorem,  we will  discuss the subject after testing the solution for a unit cell.

As the numerical simulation indicates that the electronic density is localized on $\pm  \bm{q}_{\mu}$,
this suggests to propose
$\lambda_{\mu}=Q_{\mu}=q^x_{\mu}+iq^y_{\mu}$
where $q^x_{\mu}$ and $q^y_{\mu}$ are the components of vector $\bm{q}_{\mu}=(q^x_{\mu},q^y_{\mu})$. Finally, the parameter $\sigma$ will be determined by imposing the ansatz to satisfy Eqns. (\ref{eq:eq1zeroA})-(\ref{eq:eq2zeroB}).

Now we test our ansatz in the zero mode equation. Using complex numbers and that $\Bar{\partial}=2\frac{\partial}{\partial z^*}$, the zero mode equation can be rewritten as,
\begin{equation}
\begin{split}
2\frac{\partial}{\partial z^*}\psi_{\pm}(z,z^*)=\mu_{\alpha}\alpha_{m}U(z,z^*)\psi_{\pm}(-z,-z^*)
  \end{split} 
  \end{equation}
where $\alpha_{m}$ indicate a magic angle and $U(z,z^*)$ is the complex form of the coupling layer potential $ U(\bm{r})$.

In the limit $\alpha\rightarrow\infty$ we can expand $U(z,z^*)$ locally around $Q_{\mu}$ (see appendix C) where the boundary layer lies, therefore,
\begin{equation}
\begin{split}
U(z,z^*)\psi_{\pm}(-z,-z^*)&\approx \frac{3z}{2}\psi_{\pm}(-z,-z^*)
  \end{split}. 
    \label{eq:Uz}
  \end{equation}
Next we use that the anti-holomorphic derivative of an analytic function is zero from where 
\begin{equation}
\begin{split}
\Bar{\partial}\psi_{\pm}(z,z^*)&=2\frac{\partial \psi_{\pm}(z,z^*)}{\partial z^*}=-\frac{z}{2\sigma^{2}}\psi_{\pm}(z,z^*)
  \end{split} 
  \label{eq:leftZERO}
  \end{equation}
Finally, we combine the left-side of the zero mode equation, Eq. (\ref{eq:leftZERO}), with the right-hand side and use Eq. 
(\ref{eq:Uz}) to obtain,
\begin{equation}
\begin{split}
-\frac{z}{\sigma^{2}}\psi_{\mp}(z,z^*)&\approx \mu_{\alpha}3\alpha_{m}z\psi_{\mp}(-z,-z^*)
  \end{split} 
  \label{eq:master}
  \end{equation}
where we see that the equation imposes the need of a symmetric or antisymmetric solution depending on the magic angle parity, given by the sign of $\mu_\alpha$. This can be numerically verified in Fig. \ref{fig:Structure_Nodalpoints} where we plot the symmetrized and antisymmetrized numerically obtained wavefunctions  for finite $\alpha$. Observe that for one of the layers $\psi_{\pm}(\bm{r})=\psi_1(\bm{r})\pm \psi_1(\bm{-r})=\psi_1(\bm{r})\mp i\mu_{\alpha}\psi_2(\bm{r})$, the other layer is obtained from $\psi_2(\bm{r})\pm \psi_2(\bm{-r})$. The resulting symmetric/antisymmetric components of $\psi_{\pm}(\bm{r})$ are purely real/imaginary respectively for odd/even $m$ (see appendix \ref{Cap:AppendixQHE}).  
Moreover, from Eq. (\ref{eq:master}) we obtain the width of the coherent Landau state,
\begin{equation}
    \lim_{m \rightarrow \infty }\sigma=\frac{1}{\sqrt{3\alpha_m}}
    \label{fig:sigma}
\end{equation}
To test these two results, in Fig. \ref{fig:GaussianLimit} we compare the evolution of the electronic density as $\alpha \rightarrow \infty$ for several magic angles, in this case for the axis $x=0$. The dashed line is the asymptotic solution  given by Eq. (\ref{eq:psi_coherentState2}) which does not contain any free parameter. 

In Fig. \ref{fig:Sigmaalpha} we show a log-log plot of $\sigma$ versus $\alpha_m$ as obtained by fitting Gaussians to the numerical results. The red line is the theoretical prediction given by Eq. (\ref{fig:sigma}) giving a very good agreement with the numerical data for higher order magic angles. In Fig. \ref{fig:Sigmaalpha} b) we also plot the maximum position of the numerically obtained wavefunctions ($|r_m|$), confirming the tendency for localization seen in the inverse participation ratio \cite{Naumis2022}. This is why the ansatz almost obey the Bloch´s theorem, i.e., zero modes are akin to other confined states in which the overlap between wave functions at different unitary cells is almost zero \cite{Naumis2021}. In fact, the set of coherent Landau levels is overcomplete \cite{BookCohen}.


\section{Magic angle quantization  rule and Quantum Hall effect: squared twisted graphene hamiltonian}

Let us now prove why magic angles are quantized and the relation with the quantum harmonic oscillator. Consider the zero mode Eq. (\ref{eq:eq1zeroA}) applied to a symmetric or antisymmetric wave function. We can scale the equation by setting $\bm{r}'=\bm{r}\alpha_m$ and, for the time being, without caring for the boundary conditions, we have that,

\begin{equation}
     \bar{\partial}' \psi_{\pm}(\bm{r}') = \mu_{\alpha} U(\bm{r}'/\alpha_m) \psi_{\pm}(-\bm{r}') 
\end{equation}
which suggests that changing $\alpha_m$ is akin to scale the unitary cell. However, we also need to preserve the boundary conditions. It turns out that if,
\begin{equation}
    \alpha_m=3m
\end{equation}
the boundary conditions of the potential $U(\bm{r})$ are preserved in a bigger unitary cell since for example,  the exponentials in the definition of $U(\bm{r})$ become $e^{i\bm{q}_{1}\cdot \bm{a}_{1,2}'}=e^{i3m \bm{q}_{1}\cdot \bm{a}_{1,2}}=e^{-i3m\phi}=1$. The same situation holds for the terms with $\bm{q}_{2}$  and $\bm{q}_{3}$. Notice that the scaling by $3$ appears due to the need to traverse three unitary cells in order to pick a phase $2\pi$ in $U(\bm{r})$ as   for example $U(\bm{r}+\bm{a}_1)=e^{-i\phi}U(\bm{r})$ and $U(\bm{r}+\bm{a}_2)=e^{-i\phi}U(\bm{r})$. This procedure is akin to consider a magnetic supercell as usually done in the Quantum Hall Effect and gives a possible explanation to the numerically observed change in the effective magnetic flux between magic angles \cite{WANGG2021,CANO2021}. The scaling argument explains why magic angles $\alpha_m$ with a given parity are separated by $3$. Now if we take into account sequences with alternate parities ($\mu_{\alpha}=\pm 1$) we have,
\begin{equation}
    \alpha_{m+1}-\alpha_m=\frac{3}{2}
\end{equation}

The previous results suggests some further connections with the Quantum Hall Effect. Now consider the square Hamiltonian $H^2$ for energy zero,
\begin{equation}
\begin{split}\label{eq:H2zz}
(-\nabla^2+\alpha^2|U(\bm{r})|^2)\psi_1(\bm{r})+\alpha A^{\dagger}(\bm{r})\psi_2(\bm{r})=0
  \end{split} 
  \end{equation}
  
By expanding the operators {\it up to first order in $z$} as detailed in Appendix \ref{Cap:AppendixQHE}, we show that
$A^{\dagger}(\bm{r}) \approx -3iL_z-3i$ where $L_z=i(zp_z-z^*p^*_z)$ is the angular momentum and $p_z=p_x-ip_y$ with $p_j=-i\hbar\partial_j$ is the momentum operator. Then by using complex notation for $H^2$, the symmetry relation between layer components wave function at the $\Gamma$ point and the operators expansion {\it up to first order}, in Appendix \ref{Cap:AppendixQHE} we prove that,
\begin{equation}
\begin{split}\label{eq:H2zzcomplex}
\left(4p_{z}p^*_{z}+\left(\frac{3\alpha}{2}\right)^2|z|^2-3\alpha L_{z}\right)\psi(z,z^*)=E_0\psi(z,z^*)
  \end{split} 
  \end{equation}
where $E_0=3\alpha$ is a constant energy that is related to the ground state as we will discuss. We remark a very important fact here in the sense that the previous derivation does not need the use of the coherent states. Now we define the creation/annihilation operator associated,  
\begin{equation}
\begin{split}\label{eq:az_operator}
a_{z}=\sqrt{\frac{3\alpha}{4\hbar}}z+i\frac{2}{\sqrt{3\alpha\hbar}}p^*_{z}
  \end{split} 
  \end{equation}
\begin{equation}
\begin{split}\label{eq:azdagger_operator}
a^{\dagger}_{z}=\sqrt{\frac{3\alpha}{4\hbar}}z^*-i\frac{2}{\sqrt{3\alpha\hbar}}p_{z}
  \end{split} 
  \end{equation}
from where we obtain an effective two-dimensional quantum harmonic oscillator Hamiltonian,
\begin{equation}
\begin{split}\label{eq:Hzz_Reduced2}
H_{zz^*}\psi(z,z^*)=\left(\hbar \omega a^{\dagger}_{z}a_{z}+\omega L_{z}\right)\psi(z,z^*)
\end{split} 
\end{equation}
with $\omega=3\alpha$. Defining the conjugate operators $a_{z^*}$ and $a^{\dagger}_{z^*}$ (see appendix D), it follows that the angular momentum is, 
\begin{equation}
\begin{split}\label{eq:Lz}
L_z=\frac{\hbar}{2}(a^{\dagger}_{z^*}a_{z^*}-a^{\dagger}_{z}a_{z})
  \end{split} 
  \end{equation}
where $L^{\dagger}_z=L_z$ and therefore Eq. (\ref{eq:Hzz_Reduced2}) is rewritten as, 
\begin{equation}
\begin{split}\label{eq:Hzz_Harmonic}
H_{zz^*}\psi(z,z^*)=\hbar \omega \left (N_{zz^*}+1 \right) \psi(z,z^*)
  \end{split} 
  \end{equation}
where $N_{zz^*}=\frac{1}{2}(a^{\dagger}_{z^*}a_{z^*}+a^{\dagger}_{z}a_{z})$ in analogy to a 2D harmonic oscillator. The constant term $\hbar \omega$ is the zero-point energy of the oscillator which in this case turns out to be $E_0=3\alpha$ and comes from the second term that appears in the definition of the $A^{\dagger}$ operator. 

\begin{figure}[h!]
\includegraphics[scale=0.27]{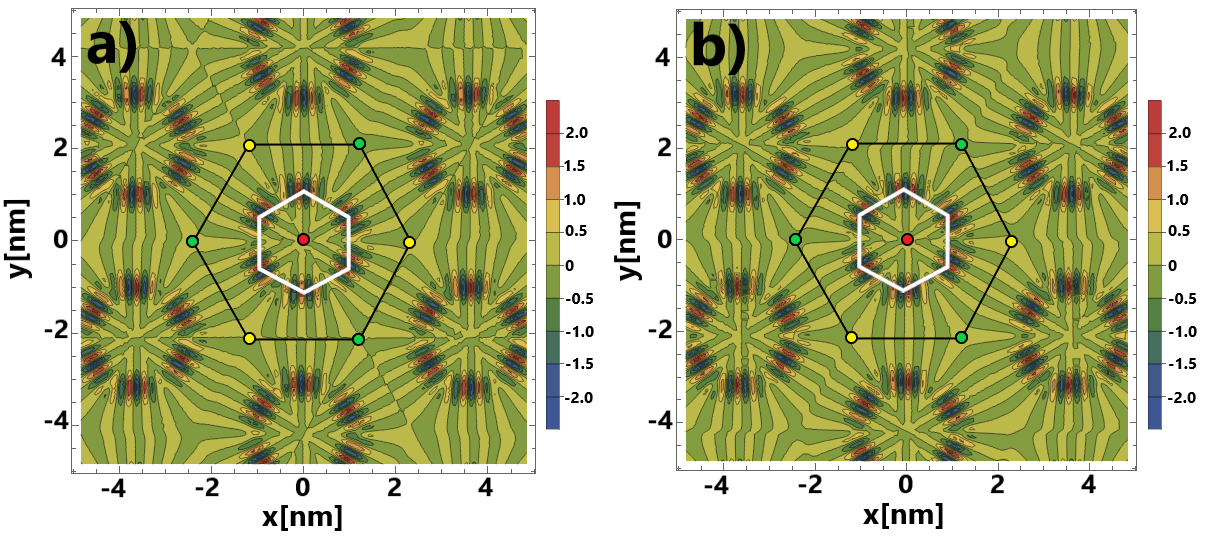}\caption{Contour plots of the a) real and b) imaginary part of the wavefunction $\psi_1(\bm{r})$ at the $m=8$ magic angle ($\alpha_8=11.345$).
The real-space unit cell is indicated as in previous figures. The white hexagon vertexes are at the points $\bm{q}_\mu$, which are the localization centers as $m \rightarrow \infty$.} \label{fig:Phase}
\end{figure}

 Thus, Eq. (\ref{eq:H2zzcomplex}) implies that $H^2$ can be identified with a quantum oscillator in which the flat band has zero quanta as $N_{zz^*}=0$. In principle, one can argue that since the effective Eq. (\ref{eq:H2zzcomplex}) is radial symmetric, $L_{z}$ conmmutes with $H_{zz^*}$ and the solutions must be all eigenfunctions of the angular momentum. There is a problem here that we discuss later on as the rotational symmetry of $H^2$ is $C_3$. But for the moment and to gain insight, for a given harmonic of the solution of Eq. (\ref{eq:H2zzcomplex}) we must have $L_{z}\psi_{m}(z,z^*)=m\psi_{m}(z,z^*)$ where the index $m$ labels a solution with a given  angular momentum. However, for the quantum harmonic oscillator $m$ only takes values $m=0,1,.., N_{zz^{*}}$. Thus $N_{zz^{*}}=0$ implies $m=0$ and we only obtain one possible state. Moreover, this condition for the angular momentum does not hold as is easy to see by looking at Fig. \ref{fig:Phase}, where we plot the imaginary and real parts of the $\Gamma$ point wavefunction for the magic angle with $m=8$.

We can clearly see a confinement in the radial direction but with nodes along a ring. The number of nodes indicates an angular momentum different from zero.  This is not a surprise as Eq. (\ref{eq:H2zzcomplex}) is a first-order expansion on $z$ and describes only a quadratic potential around the origin. How can modify this? There are many ways, as for example by using a higher order expansion of the Hamiltonian. A second way is to observe that at magic angles there is a precise energy and angular momentum balance as in the quantum harmonic oscillator. 

To understand this, first we prove in general the equipartition of energy at magic angles. Considering that $\alpha=\alpha_m$ we can apply the operator $\partial$ to the zero mode Eq. (\ref{eq:eq1zeroA}), from where,
\begin{equation}
    \partial \bar{\partial} \psi_1(\bm{r})= \alpha_{m}\mu_{\alpha}[\partial U(\bm{r}) \psi_1(-\bm{r})+U(\bm{r})\partial  \psi_1(-\bm{r})] 
\end{equation}

\begin{figure}[h!]
\includegraphics[scale=0.22]{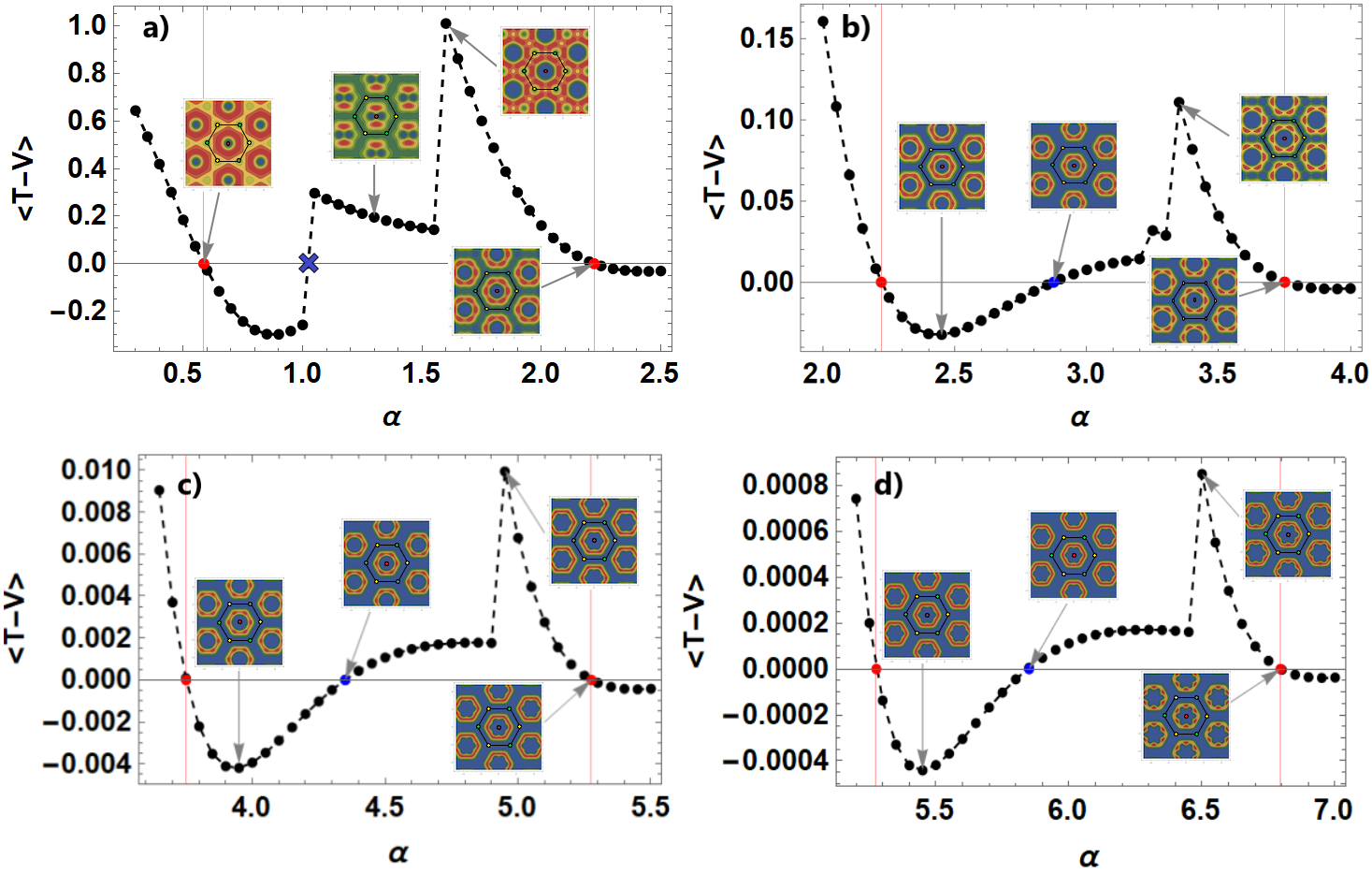}\caption{Showing the energy equipartition. The expected value $\langle T-V  \rangle$ is plotted as function of  $\alpha$ at the $\Gamma$-point for one layer. The red vertical lines indicate magic angles. Red and blue points satisfy the relation $\langle T \rangle_{m}=\langle V\rangle_{m}$ and coincide with the magic angles. The insets are the total electronic densities $\rho(\bm{r})=\rho_1(\bm{r})+\rho_2(\bm{r})$ at the indicated $\alpha$. The discontinuities arise when the upper band touches the flat band. Each panel is for a different $\alpha$ regime.} \label{fig:TV}
\end{figure}

Using $\partial  \bar{\partial}=\nabla^{2}$ and that $D^{*}(-\bm{r})(\chi_1({\bm{r}}),\chi_2(\bm{r}))^{T}=0$, as well as the symmetry relations at the $\Gamma$ point $\chi_1({\bm{r}})=\mu_{\alpha}\psi_2(\bm{r})=i (\mu_{\alpha})^{2} \psi_1(\bm{-r})$, we obtain,
\begin{equation}
     -\nabla^{2}\psi_{\pm}(\bm{r})= (\alpha_m^{2} |U(\bm{r})|^{2}+\alpha_m\partial U(r)) \psi_{\pm}(\bm{r}) 
     \label{eq:demequipart}
\end{equation}
By taking the expected values of the previous equation and using the symmetry of the potential, it  follows the equipartition of kinetic and confinement energies,
\begin{equation}
    \langle T  \rangle_{m} = -\langle \nabla^{2}  \rangle_{m}=\alpha_m^{2}\langle |U(\bm{r})|^{2} \rangle_m 
    \label{eq:finebalance}
\end{equation}
In Fig. \ref{fig:TV} we plot $\langle T-V  \rangle$ as a function of $\alpha$, where $V=\alpha^2  \langle|U(\bm{r})|^{2}\rangle$. We see that the equipartition is observed at magic angles as predicted from Eq. (\ref{eq:demequipart}).

For $m \rightarrow \infty$, the confinement is so strong that we can replace $\langle |U(\bm{r})|^{2} \rangle$ by a constant $(1/12)\sum_{\mu} (|U(\bm{q}_{\mu})|^{2}+|U(-\bm{q}_{\mu})|^{2})\approx 1$, where the $12$ in the denominator comes from the normalization with six localization centers at two layers. Therefore the expected value of the kinetic energy is,
\begin{equation}
    \langle T  \rangle_{m} = -\langle \nabla^{2}  \rangle_{m} \approx \alpha_m^{2} 
    \label{eq:kin}
\end{equation}

Thus we conclude that the kinetic and confinement energies are quantized and follow an energy equipartition as in the harmonic oscillator. Let us discuss the angular momentum. By taking the expected values in $H^{2}$, we obtain that for magic angles,
\begin{equation}
    \langle 1| A^{\dagger}(\bm{r}) |2\rangle_m +\langle 2| A(\bm{r}) |1\rangle_m=-\alpha_m 
    \label{eq:AandU2}
\end{equation}
where,
\begin{equation}
    \langle 1| A^{\dagger}(\bm{r}) |2\rangle_m =\int_{BZ} \psi_{1}^{*}(r) A^{\dagger}(\bm{r}) \psi_{2}(r) dS
\end{equation}
and 
\begin{equation}
    \langle 2| A(\bm{r}) |1\rangle_m =\int_{BZ} \psi_{2}^{*}(r) A(\bm{r}) \psi_{1}(r) dS
\end{equation}

\begin{figure}[h!]
\includegraphics[scale=0.60]{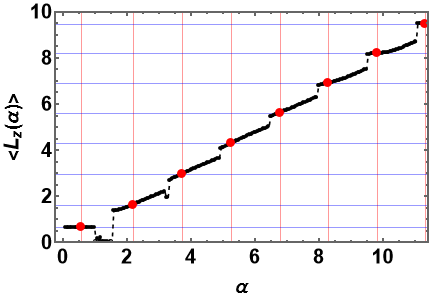}\caption{Numerically found $\langle L_{z}  \rangle_m$ as function of $\alpha$. Magic angles are indicated by the vertical lines. The horizontal lines and red points indicate the values of $\langle L_{z}  \rangle_m$ at each magic angle at the $\Gamma$-point. Observe that the distance between horizontal lines converges into one.} \label{fig:Lz_alpha}
\end{figure}

with $dS$ the surface differential and the integral is carried along the Brillouin zone. Using the quantization of $\alpha_m$ we obtain,
\begin{equation}
    \langle 1| A^{\dagger}(\bm{r}) |2\rangle_m +\langle 2| A(\bm{r}) |1\rangle_m=-3m 
    \label{eq:AandU2b}
\end{equation}
for a given parity. This proves that the eigenvalues of the off diagonal part of $H^{2}$ are $3m$. Thus the situation  is somewhat similar to  Eq. (\ref{eq:H2zzcomplex}). To see this more clearly, we use polar coordinates where $r$ is the radius and $\nu$ the polar angle.  $A^{\dagger}(\bm{r})$ is given by,
\begin{equation}
     \begin{split}
A^{\dagger}(\bm{r}) & =\sum_{\mu=1}^{3}e^{-i\bm{q}_{\mu}\cdot\bm{r}}[2(\bm{q}_{\mu}^{\perp}\cdot \bm{\hat{e}}_rP_r+\bm{q}_{\mu}^{\perp}\cdot \bm{\hat{e}}_{\nu}\frac{L_{\nu}}{r})-i]\\
  \end{split} 
  \label{eq:AGfirst}
\end{equation}
where the radial operator is $P_r=-i\partial_r$ and the angular part of the momentum is $L_{\nu}=-i\partial_{\nu}$. A similar equation is obtained for $A(\bm{r})$.
As the confinement is very strong with $r \approx 1$, and also as $\bm{q}_{\mu}^{\perp}\cdot \bm{\hat{e}}_r=0$ and $\bm{q}_{\mu}^{\perp}\cdot \bm{\hat{e}}_{\nu}=1$ at the localization centers $\pm \bm{q}_{\mu}$, we write,
\begin{equation}
     \begin{split}
A^{\dagger}(\bm{r}) & \approx A_0^{\dagger}(r)+\sum_{\mu=1}^{3}e^{-i\bm{q}_{\mu}\cdot\bm{r}} (2L_{\nu}-i)\\
  \end{split} 
  \label{eq:AGfirst1}
\end{equation}
where $A_0^{\dagger}(r)$ comes from the radial momentum contribution that we expect to be small. This is confirmed in table I, as  $\lim_{m \rightarrow \infty} \langle 1|A_0^{\dagger}|2 \rangle \approx 0.16$. The only way to be consistent with Eq. (\ref{eq:AandU2}) is to have $L_{\nu}\psi_{\pm}(\bm{r}) \approx m \psi_{\pm}(\bm{r})$.

In Fig. \ref{fig:Lz_alpha} we do see that $\langle L_z \rangle$ asymptotically grows by one on each magic angle. We comment here that  the jumps in Fig. \ref{fig:Lz_alpha} occur whenever the ground state of $H^{2}$ hybridizes with its upper neighbor band.

\begin{table}
\begin{center}
\begin{tabular}{ |p{0.3cm}|p{2.0cm}| p{2.0cm}|}

\hline
$m$ & $\langle 1|A_0^{\dagger} |2\rangle_m$ & $\Delta A_0(m)$\\
\hline
$5$ & $0.419$ & -\\
\hline
$6$ & $0.257$ & $0.162$\\
\hline
$7$ & $0.211$ &$0.046$\\
\hline
$8$ & $0.179$ & $0.032$\\
\hline
$9$ & $0.164$ & $0.015$\\
\hline
\end{tabular}
\caption{\label{table} Convergence of the operator $A_0^{\dagger}(m)$ matrix element as function of $m$. The third column is the difference between two successive contributions $\Delta(m)=\langle 1|A_0^{\dagger} |2 \rangle_m -\langle 1|A_0^{\dagger}|2  \rangle_{m-1} $.}\label{A0_constant}
\end{center}
\end{table}

There is a simple interpretation of why the previous results are akin to have in Eq. (\ref{eq:Hzz_Reduced2}) an effective $N_{zz^*}\neq 0 $ and $m \neq 0$. The confinement centers for $m \rightarrow \infty$ are not in the origin and this automatically implies an angular momentum different from zero when viewed from the origin of coordinates.


Eq. (\ref{eq:AGfirst}) can be compared with the interlayer currents between bipartite layers that were investigated in a previous work \cite{Naumis2022}. Apart from a dimensional constant, both expressions are proportional from where we can relate the interlayer currents with the angular momenta. The previous results indicate that such currents are quantized and increase with the order of $m$.

\section{Conclusions}

In conclusion, we showed that in the twisted bilayer hamiltonian, zero flat band modes converge into coherent Landau levels. The shape and dispersion of the zero-mode wavefunctions as a function of the twist angles was found and showed an excellent agreement with the numerical results. Then we proved that the squared twisted bilayer hamiltonian, up to first order, describes a quantum harmonic oscillator. Thereafter it was found that for high order magic angles, the strong confinement and the symmetry of the potential lead to solutions with well defined angular momentum. By using a scaling argument, this allows to obtain the $3/2$ magic angle quantization rule which was observed numerically \cite{Tarnpolsky2019}.  Another important consequence is that the angular momentum can be related with interlayer currents between each graphene's bipartite lattice which are thus quantized.

Let us add that in the chiral model, the interlayer tunneling in the region of AA stacking is artificially switched-off. Nevertheless, in the full description of the continuum limit for a more real twisted bilayer graphene model, there is a persistent localization of zero-energy wavefunction in the region of AA stacking \cite{MacDonald2011}. However, the region of AA stacking is much reduced than AB stacking due to lattice relaxation in real samples with small twist angles \cite{Tarnpolsky2019}. Consequently, in the limit $\alpha\rightarrow \infty$ the interlayer tunneling of AA stacking is neglected, and therefore the chiral limit is recovered. It is also possible to improve the presented study simply by taking into account the hopping between AA atoms at the individual graphene layers. This can be done in a more or less straightforward fashion by using the AA hopping as a perturbation parameter as done with random binary alloys in the band split regimen or in doped graphene \cite{Kirkpatrick1972,Naumis2002,Barrios2013}.

Nevertheless, our work does not pretend to completely solve the problem. Instead, we showed that a famous TBG model can be transformed into an effective quantum Hall effect hamiltonian once the hamiltonian is squared. This approach is akin to a supersymmetric  transformation which seems to play a role in the proposed equivalence between the squared TBG electron hamiltonian and a phonon hamiltonian for flexible systems  \cite{Naumis2021}. Flat-band modes are thus mapped into zero-frequency floppy modes produced by a lack of mechanical constraints  \cite{Huerta2002PLA,Huerta2004,Flores2010, NavarroL2021,Flores2011,Moukarzel_2022}.

Due to the multiple mathematical and physical properties of coherent states, this opens many exciting technological and physical possibilities in moir\'e materials, as for example, the possibility of controlling coherency by manipulation of the twisting angle or the self-duality property of the wave functions \cite{Hofstadter_1976}.
In particular, coherent modes and the strong electron-electron  coupling can be used in quantum computation applications \cite{Friis2015} or electron
analogues to coherent optical effects \cite{Betancur-Ocampo2019}.\\

This work was supported by UNAM DGAPA PAPIIT IN102620 (L.A.N.L. and G.G.N.) and CONACyT project 1564464. We thank Patrick Ledwith, Eslam Khalaf, Jie Wang  at Harvard University and F. Guinea, Pierre Pantaleon at IMDEA, Spain, for useful comments on this project.

\section{Appendix} 
\subsection{Wave-functions Fourier coefficients in reciprocal space}
Here we use as basis the wave vectors  $\Phi(\bm{r})=\begin{pmatrix} 
\psi_1(\bm{r}) ,
\psi_2(\bm{r}),
\chi_1(\bm{r}),
\chi_2(\bm{r})
\end{pmatrix}^T$ where the index $1,2$ represents each graphene layer and $\psi_j(\bm{r})$ and $\chi_j(\bm{r})$ are the Wannier orbitals on each inequivalent site of the graphene's unit cell. 

A general Bloch's wave function with momentum $\bm{k}$ in the mBZ at each sublattice has the form
\begin{equation}
\begin{split}\label{eq:spinor2}
\Psi_{\bm{k}}(\bm{r})=\begin{pmatrix} \psi_{\bm{k},1}(\bm{r})\\
\psi_{\bm{k},2}(\bm{r})
  \end{pmatrix} =\sum_{mn}\begin{pmatrix} a_{mn}\\
 b_{mn}e^{i\bm{q}_1\cdot\bm{r}}
  \end{pmatrix}e^{i(\bm{K}_{mn}+\bm{k})\cdot\bm{r}}
  \end{split} 
  \end{equation}
  
\begin{equation}
\begin{split}\label{eq:spinor3}
\chi_{\bm{k}}(\bm{r})=\begin{pmatrix} \chi_{\bm{k},1}(\bm{r})\\
\chi_{\bm{k},2}(\bm{r})
  \end{pmatrix} =\sum_{mn}\begin{pmatrix} c_{mn}\\
 d_{mn}e^{i\bm{q}_1\cdot\bm{r}}
  \end{pmatrix}e^{i(\bm{K}_{mn}+\bm{k})\cdot\bm{r}}
\end{split} 
\end{equation}
  
where $a_{mn}(b_{mn})$ are the Fourier coefficients of layer 1 (layer 2) for sublattice $A$ and $c_{mn}(d_{mn})$ are the Fourier coefficients of layer 1 (layer 2) for sublattice $B$, and $\bm{K}_{mn}=m\bm{b}_1+n\bm{b}_2$ with $\bm{b}_{1,2}$ are the two moiré Brillouin zone vectors. 
If we substitute Eq. (\ref{eq:spinor2}) and Eq. (\ref{eq:spinor3}) into $\mathcal{H}\Phi_{\bm{k}}(\bm{r})=E\Phi_{\bm{k}}(\bm{r})$ we can calculate the eigenfunctions of $\mathcal{H}$,

\begin{equation}
\begin{split}\label{eq:Eq1chiral}
(\bm{K}^{\prime x}_{mn}+i\bm{K}^{\prime y}_{mn})a_{mn} +\alpha(b_{mn}&+e^{i\phi}b_{m+1,n}\\
&+e^{-i\phi}b_{m,n+1})=Ea_{mn}
  \end{split} 
  \end{equation}
 \begin{equation}
\begin{split}\label{eq:Eq2chiral}
(\bm{K}^{\prime x}_{mn}+i(\bm{K}^{\prime y}_{mn} +\bm{q}_{1}))b_{mn}+&\alpha(a_{mn}+e^{i\phi}a_{m-1,n}\\
&+e^{-i\phi}b_{m,n-1})=Eb_{mn}
  \end{split} 
  \end{equation}
\begin{equation}
\begin{split}\label{eq:Eq3chiral}
(\bm{K}^{\prime x}_{mn}-i\bm{K}^{\prime y}_{mn})c_{mn} +\alpha(d_{mn}&+e^{-i\phi}d_{m+1,n}\\
&+e^{i\phi}d_{m,n+1})=Ec_{mn}
  \end{split} 
  \end{equation}
 \begin{equation}
\begin{split}\label{eq:Eq4chiral}
(\bm{K}^{\prime x}_{mn}-i(\bm{K}^{\prime y}_{mn} +\bm{q}_{1}))d_{mn}+&\alpha(c_{mn}+e^{-i\phi}c_{m-1,n}\\
&+e^{i\phi}c_{m,n-1})=Ed_{mn}
  \end{split} 
  \end{equation}
where $\bm{K}^{\prime}_{mn}=\bm{K}_{mn}+\bm{k}$.

Here Eqns. (\ref{eq:Eq1chiral}-\ref{eq:Eq4chiral}) form a coupled linear system that can be solved to find the corresponding eigenvalues. In general, there are $L=(2N+1)\times(2N+1)$ coefficients with $N$ the range of the matrix and $2N+1$ the elements in the set, therefore, the Hamiltonian matrix has dimension $D=4L$.
In a similar way we can obtain the eigenfunctions of $H^{2}$ but the system is easier write as the matrix is only of size $2\times2$. Observe that all eigenfunctions of  $\mathcal{H}$ are always eigenfunctions of $H^{2}$ but the converse is not true. This was discussed in detail elsewhere \cite{Naumis2021}.
\\ 
\\
The system can be further reduced by using the $C_3$ symmetry. We denote the corresponding rotational operators $\bm{R}_{\phi}$ and $\bm{R}_{2\phi}$ by the angle $\phi$ and $2\phi$ respectively. Their matrix representations $D_{\phi}$, $D_{2\phi}$ and the identity, have eigenvalues $w=\big\{1,e^{i\phi},e^{-i\phi}\big\}$. Eigenfunctions of the hamiltonian are also eigenfunctions of such operators, and thus we have the relation $\Psi_{\bm{k}}(\bm{R}_{2\phi}(\bm{r}))=e^{i\phi}\Psi_{\bm{k}}(\bm{r})$ and $\Psi_{\bm{k}}(\bm{R}_{\phi}(\bm{r}))=e^{-i\phi}\Psi_{\bm{k}}(\bm{r})$, with the property $\bm{R}_{\phi}(\bm{q}_{\mu})\cdot\bm{r}=\bm{q}_{\mu}\cdot\bm{R}^{-1}_{\phi}(\bm{r})$. One can obtain a relationship between coefficients using such rotations to reduce the problem into one trigonal sector.\\

To perform such calculation, it is very useful to have the moiré reciprocal basis vector transformation rules under the rotations.  We reproduce below such useful rules, 
\begin{equation}
\begin{split}\label{eq:rotate}
\bm{R}_{\phi}(\bm{b}_1)=\bm{b}_2-\bm{b}_1,
\\
\bm{R}_{\phi}(\bm{b}_2)=-\bm{b}_1,
\\
\bm{R}_{2\phi}(\bm{b}_1)=-\bm{b}_2,
\\
\bm{R}_{2\phi}(\bm{b}_1)=-\bm{b}_2+\bm{b}_1.
  \end{split} 
  \end{equation}
  
\begin{figure}[h!]
{
   \includegraphics[scale =0.5] {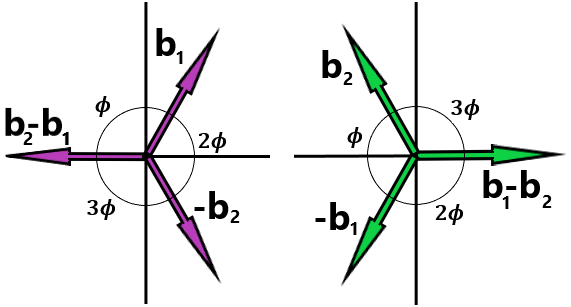}
 }
\label{fig:Rotations}
\caption{moiré vectors rotational $C_{3}$ rules. The operator $\bm{R}_{\phi}$ rotate by $\phi=\frac{2\pi}{3}$ moiré vectors $\bm{b}_{1}$ ($\bm{b}_2$) indicated by purple (green) arrows.}
\end{figure}

\subsection{Fourier coefficients in $\bm{\Gamma}$-point reciprocal}
  
In the $\bm{\Gamma}$ point,  using the symmetry of the Hamiltonian it can be proved that the spinor components are related through $\psi_{2,\bm{\Gamma}}(\bm{r})=i\mu_{\alpha}\psi_{1,\bm{\Gamma}}(-\bm{r})$. Tarnoposky et. al \cite{Tarnpolsky2019} found that, 
\begin{equation}
\begin{split}\label{eq:Gamma_zero}
\Bar{\partial}\psi_{\bm{\Gamma},1}(\bm{r})\mp\alpha U(\bm{r})\psi_{\bm{\Gamma},1}(-\bm{r})=E_{\Gamma}\psi_{\bm{\Gamma},1}(-\bm{x},\bm{y})
  \end{split} 
  \end{equation}
Following this analysis a reduced equation for the Fourier coefficients can be written as,
\begin{equation}
\begin{split}\label{eq:fourier_eq_gamma}
|\bm{K}_{mn}+\bm{q}_1|e^{i\theta(\bm{K}_{mn}+\bm{q}_1)}a_{mn}\mp\alpha(a_{mn}+e^{i\phi}a_{-m,-n+1}\\
\\ +e^{-i\phi}a_{-m+1,-n})=E_{\Gamma}a_{mn}
  \end{split} 
  \end{equation}
From the $\bm{\Gamma}=\bm{q}_{1}$ point the symmetry of the component of the spinors  $\psi_{2,\bm{\Gamma}}(\bm{r})=i\mu_{\alpha}\psi_{1,\bm{\Gamma}}(-\bm{r})$, we found the following relation,
\begin{equation}
\begin{split}
\sum_{m,n}b_{mn}e^{i(\bm{K}_{mn}+2\bm{q}_1)\cdot\bm{r}}=i\mu_{\alpha}\sum_{m^{\prime}n^{\prime}}a_{m^{\prime}n^{\prime}}e^{-i(\bm{K}_{m^{\prime}n^{\prime}}+\bm{q}_1)\cdot\bm{r}}
  \end{split} 
  \end{equation}
or simplifying,
\begin{equation}
\begin{split}\label{eq:ab_gamma_sum}
\sum_{m,n}b_{mn}e^{i\bm{K}_{mn}\cdot\bm{r}}=i\mu_{\alpha}\sum_{m^{\prime},n^{\prime}}a_{m^{\prime}n^{\prime}}e^{i(-\bm{K}_{m^{\prime}n^{\prime}}-3\bm{q}_1)\cdot\bm{r}}
  \end{split} 
  \end{equation}
note that $\bm{b}_1+\bm{b}_2=-3\bm{q}_1$ and Eq. (\ref{eq:ab_gamma_sum}) is rewritten as,
\begin{equation}
\begin{split}\label{eq:ab_gamma_sum2}
\sum_{m,n}b_{mn}e^{i\bm{K}_{mn}\cdot\bm{r}}=i\mu_{\alpha}\sum_{m^{\prime},n^{\prime}}a_{m^{\prime}n^{\prime}}e^{i\bm{K}_{1-m^{\prime},1-n^{\prime}}\cdot\bm{r}}
  \end{split} 
  \end{equation}
change index $m=1-m^{\prime}\rightarrow m^{\prime}=1-m$ and $n=1-n^{\prime}\rightarrow n^{\prime}=1-n$ in Eq. (\ref{eq:ab_gamma_sum2}), therefore,
\begin{equation}
\begin{split}\label{eq:ab_gamma_sum3}
\sum_{m,n}(b_{mn}-i\mu_{\alpha}a_{1-m,1-n})e^{i\bm{K}_{mn}\cdot\bm{r}}=0
  \end{split}
  \end{equation}
Finally, the relation between Fourier coefficients of each layer in $\bm{\Gamma}$ point is simply,
\begin{equation}
\begin{split}
b_{mn}=i\mu_{\alpha}a_{1-m,1-n}
  \end{split} 
  \end{equation}
with  $\mu_{\alpha}=\pm1$ and also follows that, 
\begin{equation}\label{eq:ImRe1}
\begin{split}
\Re{b_{mn}}=-\mu_{\alpha}\Im{a_{1-m,1-n}}
  \end{split} 
  \end{equation}
and
\begin{equation}\label{eq:ImRe2}
\begin{split}
\Im{b_{mn}}=\mu_{\alpha}\Re{a_{1-m,1-n}}
  \end{split} 
  \end{equation}
Using the $C_{3}$ symmetry of the wave-function at the $\Gamma$-point, $\bm{k}=\bm{q}_{1}$, 
\begin{equation}\label{eq:C3_wavefunction}
\begin{split}
\Psi{(\bm{R}_{\phi}(\bm{r}))}=\sum_{m,n}\begin{pmatrix}
a_{mn}\\
b_{mn}e^{i\bm{R}_{\phi}(\bm{q}_{1})\cdot\bm{r}}
  \end{pmatrix}e^{i\bm{R}_{\phi}(m\bm{b}_1+n\bm{b}_2+\bm{q}_1)\cdot\bm{r}}
  \end{split} 
  \end{equation}
where $m\bm{R}_{\phi}(\bm{b}_1)+n\bm{R}_{\phi}(\bm{b}_2)+\bm{R}_{\phi}(\bm{q}_1)=m(\bm{b}_2-\bm{b}_1)+n(-\bm{b}_2)+\bm{q}_{2}$. From the rotated vectors in Eq. (\ref{eq:C3_wavefunction}) follows that, 
\begin{equation}\label{eq:C3_wavefunction2}
\begin{split}
\sum_{m,n}a_{mn}e^{i(m\bm{b}_{1}+n\bm{b}_{2})\cdot\bm{r}}=e^{i\phi}
\sum_{m,n}a_{mn}e^{i((-m-n+1)\bm{b}_{1}+m\bm{b}_{2})\cdot\bm{r}}
  \end{split} 
  \end{equation}
using new index $m^{\prime}=-m-n+1\rightarrow n=1-m^{\prime}-n^{\prime}$ and $n^{\prime}=m$ follows that, 
\begin{equation}\label{eq:C3_wavefunction3}
\begin{split}
\sum_{m,n}a_{mn}e^{i(m\bm{b}_{1}+n\bm{b}_{2})\cdot\bm{r}}=e^{i\phi}
\sum_{m,n}a_{n^{\prime},1-m^{\prime}-n^{\prime}}e^{i(m^{\prime}\bm{b}_{1}+n^{\prime}\bm{b}_{2})\cdot\bm{r}}
  \end{split} 
  \end{equation}
therefore,
\begin{equation}\label{eq:C3_amn}
\begin{split}
e^{-i\phi}a_{mn}=
a_{n,-m-n+1}
  \end{split} 
  \end{equation}
and, 
\begin{equation}\label{eq:C3_amn2}
\begin{split}
e^{i\phi}a_{mn}=
a_{-m-n+1,m}
  \end{split}.
  \end{equation}
using this last $C_3$ symmetry in (\ref{eq:Gamma_zero}) at the $l$ magic angle it follows that, 
\begin{equation}\label{eq:EgammaReducida}
\begin{split}
(i\frac{\sqrt{3}}{2}(m-n)&+1-\frac{3}{2}(m+n))a_{m,n}\mp\alpha_{l} (a_{1-m,1-n}\\
&+a_{-n,m+n}+a_{m+n,-m})=0
  \end{split}.
  \end{equation}
where $K_{mn}=m\bm{b}_1+n\bm{b}_2=(\frac{\sqrt{3}}{2}(m-n),\frac{3}{2}(m+n)-1)$ is the moiré reciprocal vector.

\subsection{Coherent Landau level solution in the limit $\alpha\rightarrow\infty$}

Consider a coherent state of the form, 
\begin{equation}
\begin{split}\label{eq:coherent_psiAppendix}
\psi(z,z^*)&=f_{\lambda}(z)e^{-\frac{1}{4\sigma^2}|z|^2}\\
&=\frac{1}{\sigma\sqrt{2\pi}}e^{\frac{1}{2\sigma^2}\lambda^* z}e^{-\frac{1}{4\sigma^2}\lambda\lambda^*}e^{-\frac{1}{4\sigma^2}|z|^2}
  \end{split}
  \end{equation}
The complete form of wave function is obtained by summing over the contributions for $\lambda$ and $-\lambda$, 
\begin{equation}
\begin{split}\label{eq:psi_coherentStateAppendix}
\psi_{\mp}(z,z^*)&=\frac{1}{\sqrt{2}}(f_{\lambda}(z)e^{-\frac{1}{4\sigma^2}|z|^2}\mp f_{-\lambda}(z)e^{-\frac{1}{4\sigma^2}|z|^2})\\
&=\frac{1}{\sigma\sqrt{2}\sqrt{2\pi}}(e^{\frac{1}{2\sigma^2}\lambda^* z-\frac{1}{4\sigma^2}\lambda\lambda^*-\frac{1}{4\sigma^2}|z|^2}\\
&\mp  e^{-\frac{1}{2\sigma^2}\lambda^* z-\frac{1}{4\sigma^2}\lambda\lambda^*-\frac{1}{4\sigma^2}|z|^2})\\
 \end{split}
 \end{equation}
where $\lambda$ is the center of the Gaussians. Is important to note that Eq. (\ref{eq:psi_coherentStateAppendix}), satisfies the relation $\psi_{\mp}(-z,-z^*)=\mp\psi_{\mp}(z,z^*)$ as we consider here the wavefunction as a sum of symmetrized/antisymmetrized functions that can be treated separately. We also have, 
\begin{equation}
\begin{split}\label{eq:psi2_coherentState}
|\psi_{\mp}(z,z^*)&|^2 \approx \frac{1}{2\sigma^2(2\pi)}(e^{-\frac{1}{2\sigma^2}|z-\lambda|^2}+ e^{-\frac{1}{2\sigma^2}|z+\lambda|^2})
 \end{split}
 \end{equation}
where the overlap term between Gaussian's is neglected in the limit $\alpha\rightarrow\infty$. On the other hand, the coupling potential can be written as, 
\begin{equation}
\begin{split}\label{eq:Ucomplex}
U(z,z^*)=-2\frac{\partial}{\partial z^*}S(z,z^*)
 \end{split}
 \end{equation}
where $\Bar{\partial}=2\frac{\partial}{\partial z^*}$ and,
\begin{equation}
\begin{split}\label{eq:Scomplex}
S(z,z^*)=\sum^3_{\mu=1}e^{-i(Q^{*}_{\mu}z+Q_{\mu}z^*)/2}
 \end{split}
 \end{equation}
where $Q_{\mu}=q^x_{\mu}+iq^y_{\mu}$. Therefore, 
\begin{equation}
\begin{split}\label{eq:ScomplexDerivative}
\frac{\partial}{\partial z^*}S(z,z^* ) &=\frac{1}{2}\sum^3_{\mu=1}(-iQ_{\mu})e^{-i(Q^{*}_{\mu}z+Q_{\mu}z^*)/2}
 \end{split}
 \end{equation}
 
Finally, $U(z,z^*)$ is the complex form of the coupling layer potential,
\begin{equation}
    U(\bm{r})=U(z,z^*)=\sum^3_{\mu=1}(iQ_{\mu})e^{-i(Q^{*}_{\mu}z+Q_{\mu}z^*)/2}
    \label{eq:Ucompledef}
\end{equation} 
Notice that such result can be obtained  straightforward from the definition of $U(\bm{r})$ and $z$ and $z^*$, yet is illustrative to use the function $S(z,z^* )$ as this quantity appears in several commutators \cite{Naumis2021}.
 
Substituting Eqns. (\ref{eq:Ucomplex})-(\ref{eq:ScomplexDerivative}) in Eq. (\ref{eq:Gamma_zero}) for magic angles, 
\begin{equation}
\begin{split}\label{eq:GammaCOmplex}
&2\frac{\partial}{\partial z^*}\psi_{\mp}(z,z^*) = \pm\alpha_{m}(-2\frac{\partial}{\partial z^*}S(z,z^*))\psi_{\mp}(-z,-z^*)\\
&=\pm\alpha_{m}(\sum^3_{\mu=1}(iQ_{\mu})e^{-i(Q^{*}_{\mu}z+Q_{\mu}z^*)/2})\psi_{\mp}(-z,-z^*)
 \end{split}
 \end{equation}
using the fact that in the limit $\alpha\rightarrow\infty$ the wave function is localized in $Q_{\mu}$ and in other points is zero, we can expand (\ref{eq:GammaCOmplex})  up to first order, 
\begin{equation}
\begin{split}\label{eq:GammaExpanded}
2\frac{\partial}{\partial z^*}\psi_{\mp}(z,z^*)
&\approx \pm\alpha_{m}\sum^3_{\mu=1}(iQ_{\mu})(1-i\frac{1}{2}(Q^{*}_{\mu}z\\
&+Q_{\mu}z^*))\psi_{\mp}(-z,-z^*)
 \end{split}
 \end{equation}
and since $\sum_{\mu}Q_{\mu}=\sum_{\mu}Q^2_{\mu}=0$ and $\sum_{\mu}|Q_{\mu}|^2=3$, it follows that, 
\begin{equation}
\begin{split}\label{eq:psiaprox}
2\frac{\partial}{\partial z^*}\psi_{\mp}(z,z^*)
  \doteq   \left(\frac{3\alpha_{m}}{2}\right)(-z)(\psi_{\mp}(z,z^*))
 \end{split}
 \end{equation}
since $\mp\psi_{\mp}(-z,-z^*)=\psi_{\mp}(z,z^*)$.
Normalizing, the final form of the ansatz wave function for $\alpha\rightarrow\infty$ is, 
\begin{equation}
\begin{split}\label{eq:ansatz}
\psi_{\pm}(z,z^*)&=\frac{1}{\sqrt{6}}\sqrt{\frac{3\alpha}{2\pi}}\sum^3_{\mu=1}(f_{Q_{\mu}}(z)e^{-\frac{1}{4\sigma^2}|z|^2}\\
&\pm  f_{-Q_{\mu}}(z)e^{-\frac{1}{4\sigma^2}|z|^2})
 \end{split}
 \end{equation}

 \subsection{Equivalence between $H^2$ and the Quantum Harmonic Oscillator}
 \label{Cap:AppendixQHE}
From the square Hamiltonian at the zero flat band $H^2\Psi(\bm{r})=0$ it follows that,
\begin{equation}
\begin{split}\label{eq:H2zzA}
(-\nabla^2+\alpha^2|U(-\bm{r})|^2)\psi_1(\bm{r})+\alpha A^{\dagger}(\bm{r})\psi_2(\bm{r})=0
  \end{split} 
  \end{equation}
which can be rewritten in complex notation and by expanding $|U(-\bm{r})|^2$ in the boundary layer, 
\begin{equation}
\begin{split}\label{eq:H2zzC}
(4p_{z}p^*_{z}+\left(\frac{3\alpha}{2}\right)^2|z|^2)\psi(z,z^*)
+i\mu_{\alpha}\alpha A^{\dagger}(z,z^*)\\
\times\psi(-z,-z^*)=0
  \end{split} 
  \end{equation}
where $A^{\dagger}(\bm{r})=A_g^{\dagger}(\bm{r})+A_f^{\dagger}(\bm{r})$, with the definitions  \cite{Naumis2022},
\begin{equation}A_g^{\dagger}(\bm{r})=-2i\sum^3_{\mu=1}e^{-i\bm{q}_{\mu}\cdot\bm{r}}\bm{q}^{\perp}_{\mu}\cdot\bm{\nabla}
\end{equation}
and 
\begin{equation}
A_f^{\dagger}(\bm{r})=-i\sum^3_{\mu=1}e^{-i\bm{q}_{\mu}\cdot\bm{r}}.
\end{equation}
We work first with $A_g^{\dagger}(\bm{r})$
\begin{equation}
\begin{split}\label{eq:Acomplex}
A_g^{\dagger}(z,z^*)\psi(-z,-z^*)&=-2i\sum^3_{\mu=1}e^{-\frac{i}{2}(Q^*_{\mu}z+Q_{\mu}z^*)}(Q^{\perp,*}_{\mu}\partial_{z^*}\\
&+Q^{\perp}_{\mu}\partial_{z})\psi(-z,-z^*)
  \end{split} 
  \end{equation}
Now for $\alpha\rightarrow\infty$, we expand the exponential up to first order and considering that $Q^\perp_{\mu}=iQ_{\mu}$, it follows that
\begin{equation}
\begin{split}\label{eq:Acomplexaprox}
A_g^{\dagger}(z,z^*)\psi(-z,-z^*)&=-2i\sum^3_{\mu=1}(1-\frac{i}{2}(Q^*_{\mu}z+Q_{\mu}z^*))\\ \times(iQ_{\mu}\partial_{z}
&-iQ_{\mu}^*\partial_{z^*})\psi(-z,-z^*)
  \end{split} 
  \end{equation}
with $\sum^3_{\mu=1}Q_{\mu}=0$ and  $\sum^3_{\mu=1}|Q_{\mu}|^2=3$ we have that, 
\begin{equation}
\begin{split}\label{eq:Az}
A_g^{\dagger}(z,z^*)\psi(-z,-z^*)&=-3i(z\partial_{z}-z^*\partial_{z^*})\psi(-z,-z^*)\\
&=3(z p_{z}-z^* p_{z^*})\psi(-z,-z^*)
  \end{split} 
  \end{equation}
Making a similar procedure for $A_f^{\dagger}(z,z^*)$ we obtain,
\begin{equation}
A_f^{\dagger}(z,z^*)\psi(-z,-z^*) \approx -3i\psi(-z,-z^*) 
\label{eq:Ag}
\end{equation}
Substituting Eq. (\ref{eq:Az}) and Eq. (\ref{eq:Ag}) in Eq. (\ref{eq:H2zzC}), 
\begin{equation}
\begin{split}\label{eq:Hzzred}
(4p_{z}p^*_{z}+&\left(\frac{3\alpha}{2}\right)^2|z|^2)\psi(z,z^*)+3i\mu_{\alpha}\alpha(z p_{z}\\
&-z^* p_{z^*})\psi(-z,-z^*)=-\mu_{\alpha}3\alpha \psi(-z,-z^*)
  \end{split} 
  \end{equation}
However the angular momentum in complex notation is defined as $L_z=i(z p_{z}-z^*p_{z^*})$, and defining the zero-point energy constant $E_0=3\alpha$, we have that,
\begin{equation}
\begin{split}\label{eq:H2complexA}
(4p_{z}p^*_{z}+\left(\frac{3\alpha}{2}\right)^2|z|^2)\psi(z,z^*)+3\alpha L_z\mu_{\alpha}& \psi(-z,-z^*)\\
=-\mu_{\alpha}E_0 \psi(-z,-z^*)
  \end{split} 
  \end{equation}
for odd parity $\mu_{\alpha}=1$ the solution is anti-symmetric $\psi(-z,-z^*)=-\psi(z,z^*)$ and for even parity $\mu_{\alpha}=-1$ the solution is symmetric $\psi(-z,-z^*)=\psi(z,z^*)$. Therefore, the equation is reduced as, 
\begin{equation}
\begin{split}\label{eq:H2Infty}
(4p_{z}p^*_{z}+\left(\frac{3\alpha}{2}\right)^2|z|^2-3\alpha L_z)\psi(z,z^*)=E_0 \psi(z,z^*)
  \end{split}   
  \end{equation}
On the other hand, the quantum harmonic oscillator in complex notation is,
\begin{equation}
\begin{split}\label{eq:Hzz_oscilatr}
H_{zz^*}\psi(z,z^*)=\left(\frac{2p_{z}p^*_{z}}{m}+\left(\frac{m\omega^2}{2}\right)|z|^2\right)\psi(z,z^*)
 \end{split} 
 \end{equation}
 from where we identified by comparison with Eq. (\ref{eq:H2Infty}) that $m=\frac{1}{2}$ and $\omega=3\alpha$, therefore we can define a creation/annihilation operators associated as, 
 \begin{equation}
\begin{split}\label{eq:az_operatorA}
a_{z}=\sqrt{\frac{3\alpha}{4\hbar}}z+i\frac{2}{\sqrt{3\alpha\hbar}}p^*_{z}
  \end{split} 
  \end{equation}
\begin{equation}
\begin{split}\label{eq:azdagger_operatorA}
a^{\dagger}_{z}=\sqrt{\frac{3\alpha}{4\hbar}}z^*-i\frac{2}{\sqrt{3\alpha\hbar}}p_{z}
  \end{split} 
  \end{equation}
from where,
\begin{equation}
\begin{split}\label{eq:azdagaz_App}
a^{\dagger}_{z}a_{z}=\frac{3\alpha}{4\hbar}|z|^2+\frac{4p_{z}p^*_{z}}{3\alpha\hbar}+i\frac{2}{\sqrt{3\alpha\hbar}}\sqrt{\frac{3\alpha}{4\hbar}}(z^*p^*_{z}-zp_{z})
  \end{split} 
  \end{equation}
multiplying both sides of Eq. (\ref{eq:azdagaz_App}) by $3\alpha\hbar$ follows that 
\begin{equation}
\begin{split}\label{eq:azdagaz_AppR}
\hbar(3\alpha)a^{\dagger}_{z}a_{z}&=\left(\frac{3\alpha}{2}\right)^2|z|^2+4p_{z}p^*_{z}-3\alpha L_z\\
&=H_{zz^{*}}-3\alpha L_z
  \end{split} 
  \end{equation}
from where we obtain an effective two-dimensional quantum harmonic oscillator Hamiltonian,
\begin{equation}
\begin{split}\label{eq:Hzz_Reduced}
H_{zz^*}\psi(z,z^*)=\left(\hbar \omega a^{\dagger}_{z}a_{z}+\omega L_{z}\right)\psi(z,z^*)
  \end{split} 
  \end{equation}
We can also define the conjugate operators,
\begin{equation}
\begin{split}\label{eq:az_operatorC}
a_{z^*}=\sqrt{\frac{3\alpha}{4\hbar}}z^*+i\frac{2}{\sqrt{3\alpha\hbar}}p_{z}
  \end{split} 
  \end{equation}
\begin{equation}
\begin{split}\label{eq:azdagger_operatorC}
a^{\dagger}_{z^*}=\sqrt{\frac{3\alpha}{4\hbar}}z-i\frac{2}{\sqrt{3\alpha\hbar}}p^*_{z}
  \end{split} 
  \end{equation}
and it follows that the angular momentum is, 
\begin{equation}
\begin{split}\label{eq:Lz2}
L_z=\frac{\hbar}{2}(a^{\dagger}_{z^*}a_{z^*}-a^{\dagger}_{z}a_{z})
  \end{split} 
  \end{equation}
where $L^{\dagger}_z=L_z$ is Hermitian and therefore Eq. (\ref{eq:Hzz_Reduced}) is rewritten as, 
\begin{equation}
\begin{split}\label{eq:Hzz_Harmonic2}
H_{zz^*}\psi(z,z^*)=\hbar \omega \left (N_{zz^*}+1 \right) \psi(z,z^*)
  \end{split} 
  \end{equation}
with $N_{zz^*}=\frac{1}{2}(a^{\dagger}_{z^*}a_{z^*}+a^{\dagger}_{z}a_{z})=\frac{1}{2}(N_{z}+N_{z^*})$ with $N_{z}=a^{\dagger}_{z}a_{z}$ and $N_{z^*}=a^{\dagger}_{z^*}a_{z^*}$ in analogy to a 2D harmonic oscillator where $\omega=3\alpha$. The constant term $\hbar \omega$ is the zero-point energy of the oscillator which in this case is exactly $E_0=3\alpha$ and comes from the $A^{\dagger}_f$ operator.

\begin{thebibliography}{62}%
\makeatletter
\providecommand \@ifxundefined [1]{%
 \@ifx{#1\undefined}
}%
\providecommand \@ifnum [1]{%
 \ifnum #1\expandafter \@firstoftwo
 \else \expandafter \@secondoftwo
 \fi
}%
\providecommand \@ifx [1]{%
 \ifx #1\expandafter \@firstoftwo
 \else \expandafter \@secondoftwo
 \fi
}%
\providecommand \natexlab [1]{#1}%
\providecommand \enquote  [1]{``#1''}%
\providecommand \bibnamefont  [1]{#1}%
\providecommand \bibfnamefont [1]{#1}%
\providecommand \citenamefont [1]{#1}%
\providecommand \href@noop [0]{\@secondoftwo}%
\providecommand \href [0]{\begingroup \@sanitize@url \@href}%
\providecommand \@href[1]{\@@startlink{#1}\@@href}%
\providecommand \@@href[1]{\endgroup#1\@@endlink}%
\providecommand \@sanitize@url [0]{\catcode `\\12\catcode `\$12\catcode
  `\&12\catcode `\#12\catcode `\^12\catcode `\_12\catcode `\%12\relax}%
\providecommand \@@startlink[1]{}%
\providecommand \@@endlink[0]{}%
\providecommand \url  [0]{\begingroup\@sanitize@url \@url }%
\providecommand \@url [1]{\endgroup\@href {#1}{\urlprefix }}%
\providecommand \urlprefix  [0]{URL }%
\providecommand \Eprint [0]{\href }%
\providecommand \doibase [0]{https://doi.org/}%
\providecommand \selectlanguage [0]{\@gobble}%
\providecommand \bibinfo  [0]{\@secondoftwo}%
\providecommand \bibfield  [0]{\@secondoftwo}%
\providecommand \translation [1]{[#1]}%
\providecommand \BibitemOpen [0]{}%
\providecommand \bibitemStop [0]{}%
\providecommand \bibitemNoStop [0]{.\EOS\space}%
\providecommand \EOS [0]{\spacefactor3000\relax}%
\providecommand \BibitemShut  [1]{\csname bibitem#1\endcsname}%
\let\auto@bib@innerbib\@empty
\bibitem [{\citenamefont {Cao}\ \emph {et~al.}(2018)\citenamefont {Cao},
  \citenamefont {Fatemi}, \citenamefont {Fang}, \citenamefont {Watanabe},
  \citenamefont {Taniguchi}, \citenamefont {Kaxiras},\ and\ \citenamefont
  {Jarillo-Herrero}}]{Cao2018}%
  \BibitemOpen
  \bibfield  {author} {\bibinfo {author} {\bibfnamefont {Y.}~\bibnamefont
  {Cao}}, \bibinfo {author} {\bibfnamefont {V.}~\bibnamefont {Fatemi}},
  \bibinfo {author} {\bibfnamefont {S.}~\bibnamefont {Fang}}, \bibinfo {author}
  {\bibfnamefont {K.}~\bibnamefont {Watanabe}}, \bibinfo {author}
  {\bibfnamefont {T.}~\bibnamefont {Taniguchi}}, \bibinfo {author}
  {\bibfnamefont {E.}~\bibnamefont {Kaxiras}},\ and\ \bibinfo {author}
  {\bibfnamefont {P.}~\bibnamefont {Jarillo-Herrero}},\ }\href
  {https://doi.org/10.1038/nature26160} {\bibfield  {journal} {\bibinfo
  {journal} {Nature}\ }\textbf {\bibinfo {volume} {556}},\ \bibinfo {pages}
  {43} (\bibinfo {year} {2018})}\BibitemShut {NoStop}%
\bibitem [{\citenamefont {Park}\ \emph {et~al.}(2021)\citenamefont {Park},
  \citenamefont {Cao}, \citenamefont {Watanabe}, \citenamefont {Taniguchi},\
  and\ \citenamefont {Jarillo-Herrero}}]{park2021}%
  \BibitemOpen
  \bibfield  {author} {\bibinfo {author} {\bibfnamefont {J.~M.}\ \bibnamefont
  {Park}}, \bibinfo {author} {\bibfnamefont {Y.}~\bibnamefont {Cao}}, \bibinfo
  {author} {\bibfnamefont {K.}~\bibnamefont {Watanabe}}, \bibinfo {author}
  {\bibfnamefont {T.}~\bibnamefont {Taniguchi}},\ and\ \bibinfo {author}
  {\bibfnamefont {P.}~\bibnamefont {Jarillo-Herrero}},\ }\href
  {https://doi.org/10.1038/s41586-021-03192-0} {\bibfield  {journal} {\bibinfo
  {journal} {Nature}\ }\textbf {\bibinfo {volume} {590}},\ \bibinfo {pages}
  {249} (\bibinfo {year} {2021})}\BibitemShut {NoStop}%
\bibitem [{\citenamefont {{Shen}}\ \emph {et~al.}(2022)\citenamefont {{Shen}},
  \citenamefont {{Ledwith}}, \citenamefont {{Watanabe}}, \citenamefont
  {{Taniguchi}}, \citenamefont {{Khalaf}}, \citenamefont {{Vishwanath}},\ and\
  \citenamefont {{Efetov}}}]{2022Ashvin}%
  \BibitemOpen
  \bibfield  {author} {\bibinfo {author} {\bibfnamefont {C.}~\bibnamefont
  {{Shen}}}, \bibinfo {author} {\bibfnamefont {P.~J.}\ \bibnamefont
  {{Ledwith}}}, \bibinfo {author} {\bibfnamefont {K.}~\bibnamefont
  {{Watanabe}}}, \bibinfo {author} {\bibfnamefont {T.}~\bibnamefont
  {{Taniguchi}}}, \bibinfo {author} {\bibfnamefont {E.}~\bibnamefont
  {{Khalaf}}}, \bibinfo {author} {\bibfnamefont {A.}~\bibnamefont
  {{Vishwanath}}},\ and\ \bibinfo {author} {\bibfnamefont {D.~K.}\ \bibnamefont
  {{Efetov}}},\ }\href@noop {} {\bibfield  {journal} {\bibinfo  {journal}
  {arXiv e-prints}\ ,\ \bibinfo {eid} {arXiv:2204.07244}} (\bibinfo {year}
  {2022})},\ \Eprint {https://arxiv.org/abs/2204.07244} {arXiv:2204.07244
  [cond-mat.mes-hall]} \BibitemShut {NoStop}%
\bibitem [{\citenamefont {Song}\ and\ \citenamefont
  {Bernevig}(2022)}]{Bernevig2022}%
  \BibitemOpen
  \bibfield  {author} {\bibinfo {author} {\bibfnamefont {Z.-D.}\ \bibnamefont
  {Song}}\ and\ \bibinfo {author} {\bibfnamefont {B.~A.}\ \bibnamefont
  {Bernevig}},\ }\href {https://doi.org/10.1103/PhysRevLett.129.047601}
  {\bibfield  {journal} {\bibinfo  {journal} {Phys. Rev. Lett.}\ }\textbf
  {\bibinfo {volume} {129}},\ \bibinfo {pages} {047601} (\bibinfo {year}
  {2022})}\BibitemShut {NoStop}%
\bibitem [{\citenamefont {{Goodwin}}\ \emph {et~al.}(2019)\citenamefont
  {{Goodwin}}, \citenamefont {{Corsetti}}, \citenamefont {{Mostofi}},\ and\
  \citenamefont {{Lischner}}}]{Zachary2019_2}%
  \BibitemOpen
  \bibfield  {author} {\bibinfo {author} {\bibfnamefont {Z.~A.~H.}\
  \bibnamefont {{Goodwin}}}, \bibinfo {author} {\bibfnamefont {F.}~\bibnamefont
  {{Corsetti}}}, \bibinfo {author} {\bibfnamefont {A.~A.}\ \bibnamefont
  {{Mostofi}}},\ and\ \bibinfo {author} {\bibfnamefont {J.}~\bibnamefont
  {{Lischner}}},\ }\href
  {https://doi.org/https://doi.org/10.1103/PhysRevB.100.121106} {\bibfield
  {journal} {\bibinfo  {journal} {Phys. Rev. B}\ }\textbf {\bibinfo {volume}
  {100}},\ \bibinfo {pages} {121106} (\bibinfo {year} {2019})}\BibitemShut
  {NoStop}%
\bibitem [{\citenamefont {Hridis}\ \emph {et~al.}(2019)\citenamefont {Hridis},
  \citenamefont {Pal}, \citenamefont {Stephen}, \citenamefont {Spitz},
  \citenamefont {Markus},\ and\ \citenamefont {Kindermann}}]{Hridis2019}%
  \BibitemOpen
  \bibfield  {author} {\bibinfo {author} {\bibnamefont {Hridis}}, \bibinfo
  {author} {\bibfnamefont {K.}~\bibnamefont {Pal}}, \bibinfo {author}
  {\bibnamefont {Stephen}}, \bibinfo {author} {\bibnamefont {Spitz}}, \bibinfo
  {author} {\bibnamefont {Markus}},\ and\ \bibinfo {author} {\bibnamefont
  {Kindermann}},\ }\href {https://doi.org/10.1103/PhysRevLett.123.186402}
  {\bibfield  {journal} {\bibinfo  {journal} {Phys. Rev. Lett.}\ }\textbf
  {\bibinfo {volume} {123}},\ \bibinfo {pages} {186402} (\bibinfo {year}
  {2019})}\BibitemShut {NoStop}%
\bibitem [{\citenamefont {Kerelsky}\ \emph {et~al.}(2019)\citenamefont
  {Kerelsky}, \citenamefont {McGilly},\ and\ \citenamefont
  {Kennes}}]{Kerelsky2019}%
  \BibitemOpen
  \bibfield  {author} {\bibinfo {author} {\bibfnamefont {A.}~\bibnamefont
  {Kerelsky}}, \bibinfo {author} {\bibfnamefont {L.}~\bibnamefont {McGilly}},\
  and\ \bibinfo {author} {\bibfnamefont {D.~e.~a.}\ \bibnamefont {Kennes}},\
  }\href {https://doi.org/10.1038/s41586-019-1431-9} {\bibfield  {journal}
  {\bibinfo  {journal} {Nature}\ }\textbf {\bibinfo {volume} {572}},\ \bibinfo
  {pages} {95} (\bibinfo {year} {2019})}\BibitemShut {NoStop}%
\bibitem [{\citenamefont {{Wang}}\ \emph
  {et~al.}(2020{\natexlab{a}})\citenamefont {{Wang}}, \citenamefont {{Yuan}},\
  and\ \citenamefont {{Fu}}}]{Fu2020}%
  \BibitemOpen
  \bibfield  {author} {\bibinfo {author} {\bibfnamefont {T.}~\bibnamefont
  {{Wang}}}, \bibinfo {author} {\bibfnamefont {N.~F.~Q.}\ \bibnamefont
  {{Yuan}}},\ and\ \bibinfo {author} {\bibfnamefont {L.}~\bibnamefont {{Fu}}},\
  }\href@noop {} {\bibfield  {journal} {\bibinfo  {journal} {arXiv e-prints}\
  ,\ \bibinfo {eid} {arXiv:2010.09753}} (\bibinfo {year}
  {2020}{\natexlab{a}})},\ \Eprint {https://arxiv.org/abs/2010.09753}
  {arXiv:2010.09753 [cond-mat.supr-con]} \BibitemShut {NoStop}%
\bibitem [{\citenamefont {Ochoa}\ and\ \citenamefont
  {Fernandes}(2021)}]{ochoa2021degradation}%
  \BibitemOpen
  \bibfield  {author} {\bibinfo {author} {\bibfnamefont {H.}~\bibnamefont
  {Ochoa}}\ and\ \bibinfo {author} {\bibfnamefont {R.~M.}\ \bibnamefont
  {Fernandes}},\ }\href@noop {} {\bibinfo {title} {Degradation of phonons in
  disordered moir\'e superlattices}} (\bibinfo {year} {2021}),\ \Eprint
  {https://arxiv.org/abs/2108.10342} {arXiv:2108.10342 [cond-mat.mes-hall]}
  \BibitemShut {NoStop}%
\bibitem [{\citenamefont {Koshino}\ and\ \citenamefont {Nam}(2020)}]{Namm2020}%
  \BibitemOpen
  \bibfield  {author} {\bibinfo {author} {\bibfnamefont {M.}~\bibnamefont
  {Koshino}}\ and\ \bibinfo {author} {\bibfnamefont {N.~N.~T.}\ \bibnamefont
  {Nam}},\ }\href {https://doi.org/10.1103/PhysRevB.101.195425} {\bibfield
  {journal} {\bibinfo  {journal} {Phys. Rev. B}\ }\textbf {\bibinfo {volume}
  {101}},\ \bibinfo {pages} {195425} (\bibinfo {year} {2020})}\BibitemShut
  {NoStop}%
\bibitem [{\citenamefont {Oka}\ and\ \citenamefont {Koshino}(2021)}]{Oka2021}%
  \BibitemOpen
  \bibfield  {author} {\bibinfo {author} {\bibfnamefont {H.}~\bibnamefont
  {Oka}}\ and\ \bibinfo {author} {\bibfnamefont {M.}~\bibnamefont {Koshino}},\
  }\href {https://doi.org/10.1103/PhysRevB.104.035306} {\bibfield  {journal}
  {\bibinfo  {journal} {Phys. Rev. B}\ }\textbf {\bibinfo {volume} {104}},\
  \bibinfo {pages} {035306} (\bibinfo {year} {2021})}\BibitemShut {NoStop}%
\bibitem [{\citenamefont {Fujimoto}\ and\ \citenamefont
  {Koshino}(2021)}]{Manato2021}%
  \BibitemOpen
  \bibfield  {author} {\bibinfo {author} {\bibfnamefont {M.}~\bibnamefont
  {Fujimoto}}\ and\ \bibinfo {author} {\bibfnamefont {M.}~\bibnamefont
  {Koshino}},\ }\href {https://doi.org/10.1103/PhysRevB.103.155410} {\bibfield
  {journal} {\bibinfo  {journal} {Phys. Rev. B}\ }\textbf {\bibinfo {volume}
  {103}},\ \bibinfo {pages} {155410} (\bibinfo {year} {2021})}\BibitemShut
  {NoStop}%
\bibitem [{\citenamefont {Dante}\ \emph {et~al.}(2021)\citenamefont {Dante},
  \citenamefont {Martin}, \citenamefont {Lede}, \citenamefont {Antoine},
  \citenamefont {Andrew}, \citenamefont {James}, \citenamefont {Cory},
  \citenamefont {D.N.}, \citenamefont {Abhay},\ and\ \citenamefont
  {Angel}}]{Rubio2021}%
  \BibitemOpen
  \bibfield  {author} {\bibinfo {author} {\bibfnamefont {K.}~\bibnamefont
  {Dante}}, \bibinfo {author} {\bibfnamefont {C.}~\bibnamefont {Martin}},
  \bibinfo {author} {\bibfnamefont {X.}~\bibnamefont {Lede}}, \bibinfo {author}
  {\bibfnamefont {G.}~\bibnamefont {Antoine}}, \bibinfo {author} {\bibfnamefont
  {M.}~\bibnamefont {Andrew}}, \bibinfo {author} {\bibfnamefont
  {H.}~\bibnamefont {James}}, \bibinfo {author} {\bibfnamefont
  {D.}~\bibnamefont {Cory}}, \bibinfo {author} {\bibfnamefont {B.}~\bibnamefont
  {D.N.}}, \bibinfo {author} {\bibfnamefont {P.}~\bibnamefont {Abhay}},\ and\
  \bibinfo {author} {\bibfnamefont {R.}~\bibnamefont {Angel}},\ }\href
  {https://doi.org/https://doi.org/10.1038/s41567-020-01154-3} {\bibfield
  {journal} {\bibinfo  {journal} {Nature Physics}\ }\textbf {\bibinfo {volume}
  {17}},\ \bibinfo {pages} {155} (\bibinfo {year} {2021})}\BibitemShut
  {NoStop}%
\bibitem [{\citenamefont {Giustino}\ \emph {et~al.}(2020)\citenamefont
  {Giustino}, \citenamefont {Lee}, \citenamefont {Trier}, \citenamefont
  {Bibes}, \citenamefont {Winter}, \citenamefont {Valent{\'{\i}}},
  \citenamefont {Son}, \citenamefont {Taillefer}, \citenamefont {Heil},
  \citenamefont {Figueroa}, \citenamefont {Pla{\c{c}}ais}, \citenamefont {Wu},
  \citenamefont {Yazyev}, \citenamefont {Bakkers}, \citenamefont {Nyg{\aa}rd},
  \citenamefont {Forn-D{\'{\i}}az}, \citenamefont {Franceschi}, \citenamefont
  {McIver}, \citenamefont {Torres}, \citenamefont {Low}, \citenamefont {Kumar},
  \citenamefont {Galceran}, \citenamefont {Valenzuela}, \citenamefont
  {Costache}, \citenamefont {Manchon}, \citenamefont {Kim}, \citenamefont
  {Schleder}, \citenamefont {Fazzio},\ and\ \citenamefont
  {Roche}}]{2021Rochee}%
  \BibitemOpen
  \bibfield  {author} {\bibinfo {author} {\bibfnamefont {F.}~\bibnamefont
  {Giustino}}, \bibinfo {author} {\bibfnamefont {J.~H.}\ \bibnamefont {Lee}},
  \bibinfo {author} {\bibfnamefont {F.}~\bibnamefont {Trier}}, \bibinfo
  {author} {\bibfnamefont {M.}~\bibnamefont {Bibes}}, \bibinfo {author}
  {\bibfnamefont {S.~M.}\ \bibnamefont {Winter}}, \bibinfo {author}
  {\bibfnamefont {R.}~\bibnamefont {Valent{\'{\i}}}}, \bibinfo {author}
  {\bibfnamefont {Y.-W.}\ \bibnamefont {Son}}, \bibinfo {author} {\bibfnamefont
  {L.}~\bibnamefont {Taillefer}}, \bibinfo {author} {\bibfnamefont
  {C.}~\bibnamefont {Heil}}, \bibinfo {author} {\bibfnamefont {A.~I.}\
  \bibnamefont {Figueroa}}, \bibinfo {author} {\bibfnamefont {B.}~\bibnamefont
  {Pla{\c{c}}ais}}, \bibinfo {author} {\bibfnamefont {Q.}~\bibnamefont {Wu}},
  \bibinfo {author} {\bibfnamefont {O.~V.}\ \bibnamefont {Yazyev}}, \bibinfo
  {author} {\bibfnamefont {E.~P. A.~M.}\ \bibnamefont {Bakkers}}, \bibinfo
  {author} {\bibfnamefont {J.}~\bibnamefont {Nyg{\aa}rd}}, \bibinfo {author}
  {\bibfnamefont {P.}~\bibnamefont {Forn-D{\'{\i}}az}}, \bibinfo {author}
  {\bibfnamefont {S.~D.}\ \bibnamefont {Franceschi}}, \bibinfo {author}
  {\bibfnamefont {J.~W.}\ \bibnamefont {McIver}}, \bibinfo {author}
  {\bibfnamefont {L.~E. F.~F.}\ \bibnamefont {Torres}}, \bibinfo {author}
  {\bibfnamefont {T.}~\bibnamefont {Low}}, \bibinfo {author} {\bibfnamefont
  {A.}~\bibnamefont {Kumar}}, \bibinfo {author} {\bibfnamefont
  {R.}~\bibnamefont {Galceran}}, \bibinfo {author} {\bibfnamefont {S.~O.}\
  \bibnamefont {Valenzuela}}, \bibinfo {author} {\bibfnamefont {M.~V.}\
  \bibnamefont {Costache}}, \bibinfo {author} {\bibfnamefont {A.}~\bibnamefont
  {Manchon}}, \bibinfo {author} {\bibfnamefont {E.-A.}\ \bibnamefont {Kim}},
  \bibinfo {author} {\bibfnamefont {G.~R.}\ \bibnamefont {Schleder}}, \bibinfo
  {author} {\bibfnamefont {A.}~\bibnamefont {Fazzio}},\ and\ \bibinfo {author}
  {\bibfnamefont {S.}~\bibnamefont {Roche}},\ }\href
  {https://doi.org/10.1088/2515-7639/abb74e} {\bibfield  {journal} {\bibinfo
  {journal} {Journal of Physics: Materials}\ }\textbf {\bibinfo {volume} {3}},\
  \bibinfo {pages} {042006} (\bibinfo {year} {2020})}\BibitemShut {NoStop}%
\bibitem [{\citenamefont {Andrews}\ and\ \citenamefont
  {Soluyanov}(2020)}]{Barth2020}%
  \BibitemOpen
  \bibfield  {author} {\bibinfo {author} {\bibfnamefont {B.}~\bibnamefont
  {Andrews}}\ and\ \bibinfo {author} {\bibfnamefont {A.}~\bibnamefont
  {Soluyanov}},\ }\href {https://doi.org/10.1103/PhysRevB.101.235312}
  {\bibfield  {journal} {\bibinfo  {journal} {Phys. Rev. B}\ }\textbf {\bibinfo
  {volume} {101}},\ \bibinfo {pages} {235312} (\bibinfo {year}
  {2020})}\BibitemShut {NoStop}%
\bibitem [{\citenamefont {Liu}\ \emph {et~al.}(2021{\natexlab{a}})\citenamefont
  {Liu}, \citenamefont {Khalaf}, \citenamefont {Lee},\ and\ \citenamefont
  {Vishwanath}}]{2021ShangL}%
  \BibitemOpen
  \bibfield  {author} {\bibinfo {author} {\bibfnamefont {S.}~\bibnamefont
  {Liu}}, \bibinfo {author} {\bibfnamefont {E.}~\bibnamefont {Khalaf}},
  \bibinfo {author} {\bibfnamefont {J.~Y.}\ \bibnamefont {Lee}},\ and\ \bibinfo
  {author} {\bibfnamefont {A.}~\bibnamefont {Vishwanath}},\ }\href
  {https://doi.org/10.1103/PhysRevResearch.3.013033} {\bibfield  {journal}
  {\bibinfo  {journal} {Phys. Rev. Research}\ }\textbf {\bibinfo {volume}
  {3}},\ \bibinfo {pages} {013033} (\bibinfo {year}
  {2021}{\natexlab{a}})}\BibitemShut {NoStop}%
\bibitem [{\citenamefont {Liu}\ \emph {et~al.}(2021{\natexlab{b}})\citenamefont
  {Liu}, \citenamefont {Khalaf}, \citenamefont {Lee},\ and\ \citenamefont
  {Vishwanath}}]{2021Liu}%
  \BibitemOpen
  \bibfield  {author} {\bibinfo {author} {\bibfnamefont {S.}~\bibnamefont
  {Liu}}, \bibinfo {author} {\bibfnamefont {E.}~\bibnamefont {Khalaf}},
  \bibinfo {author} {\bibfnamefont {J.~Y.}\ \bibnamefont {Lee}},\ and\ \bibinfo
  {author} {\bibfnamefont {A.}~\bibnamefont {Vishwanath}},\ }\href
  {https://doi.org/10.1103/PhysRevResearch.3.013033} {\bibfield  {journal}
  {\bibinfo  {journal} {Phys. Rev. Research}\ }\textbf {\bibinfo {volume}
  {3}},\ \bibinfo {pages} {013033} (\bibinfo {year}
  {2021}{\natexlab{b}})}\BibitemShut {NoStop}%
\bibitem [{\citenamefont {{Ledwith}}\ \emph {et~al.}(2022)\citenamefont
  {{Ledwith}}, \citenamefont {{Vishwanath}},\ and\ \citenamefont
  {{Parker}}}]{2022Ledwith_Vortex}%
  \BibitemOpen
  \bibfield  {author} {\bibinfo {author} {\bibfnamefont {P.~J.}\ \bibnamefont
  {{Ledwith}}}, \bibinfo {author} {\bibfnamefont {A.}~\bibnamefont
  {{Vishwanath}}},\ and\ \bibinfo {author} {\bibfnamefont {D.~E.}\ \bibnamefont
  {{Parker}}},\ }\href@noop {} {\bibfield  {journal} {\bibinfo  {journal}
  {arXiv e-prints}\ ,\ \bibinfo {eid} {arXiv:2209.15023}} (\bibinfo {year}
  {2022})},\ \Eprint {https://arxiv.org/abs/2209.15023} {arXiv:2209.15023
  [cond-mat.str-el]} \BibitemShut {NoStop}%
\bibitem [{\citenamefont {Ledwith}\ \emph {et~al.}(2022)\citenamefont
  {Ledwith}, \citenamefont {Vishwanath},\ and\ \citenamefont
  {Khalaf}}]{2022Vishwanath}%
  \BibitemOpen
  \bibfield  {author} {\bibinfo {author} {\bibfnamefont {P.~J.}\ \bibnamefont
  {Ledwith}}, \bibinfo {author} {\bibfnamefont {A.}~\bibnamefont
  {Vishwanath}},\ and\ \bibinfo {author} {\bibfnamefont {E.}~\bibnamefont
  {Khalaf}},\ }\href {https://doi.org/10.1103/PhysRevLett.128.176404}
  {\bibfield  {journal} {\bibinfo  {journal} {Phys. Rev. Lett.}\ }\textbf
  {\bibinfo {volume} {128}},\ \bibinfo {pages} {176404} (\bibinfo {year}
  {2022})}\BibitemShut {NoStop}%
\bibitem [{\citenamefont {Xu}\ and\ \citenamefont {Balents}(2018)}]{Xu2018}%
  \BibitemOpen
  \bibfield  {author} {\bibinfo {author} {\bibfnamefont {C.}~\bibnamefont
  {Xu}}\ and\ \bibinfo {author} {\bibfnamefont {L.}~\bibnamefont {Balents}},\
  }\href {https://doi.org/10.1103/PhysRevLett.121.087001} {\bibfield  {journal}
  {\bibinfo  {journal} {Phys. Rev. Lett.}\ }\textbf {\bibinfo {volume} {121}},\
  \bibinfo {pages} {087001} (\bibinfo {year} {2018})}\BibitemShut {NoStop}%
\bibitem [{\citenamefont {Wu}(2019)}]{2019FengCheng}%
  \BibitemOpen
  \bibfield  {author} {\bibinfo {author} {\bibfnamefont {F.}~\bibnamefont
  {Wu}},\ }\href {https://doi.org/10.1103/PhysRevB.99.195114} {\bibfield
  {journal} {\bibinfo  {journal} {Phys. Rev. B}\ }\textbf {\bibinfo {volume}
  {99}},\ \bibinfo {pages} {195114} (\bibinfo {year} {2019})}\BibitemShut
  {NoStop}%
\bibitem [{\citenamefont {Gonz\'alez}\ and\ \citenamefont
  {Stauber}(2020)}]{Stauber2020}%
  \BibitemOpen
  \bibfield  {author} {\bibinfo {author} {\bibfnamefont {J.}~\bibnamefont
  {Gonz\'alez}}\ and\ \bibinfo {author} {\bibfnamefont {T.}~\bibnamefont
  {Stauber}},\ }\href {https://doi.org/10.1103/PhysRevB.102.081118} {\bibfield
  {journal} {\bibinfo  {journal} {Phys. Rev. B}\ }\textbf {\bibinfo {volume}
  {102}},\ \bibinfo {pages} {081118} (\bibinfo {year} {2020})}\BibitemShut
  {NoStop}%
\bibitem [{\citenamefont {Wu}\ \emph {et~al.}(2018)\citenamefont {Wu},
  \citenamefont {MacDonald},\ and\ \citenamefont {Martin}}]{2018Wu}%
  \BibitemOpen
  \bibfield  {author} {\bibinfo {author} {\bibfnamefont {F.}~\bibnamefont
  {Wu}}, \bibinfo {author} {\bibfnamefont {A.~H.}\ \bibnamefont {MacDonald}},\
  and\ \bibinfo {author} {\bibfnamefont {I.}~\bibnamefont {Martin}},\ }\href
  {https://doi.org/10.1103/PhysRevLett.121.257001} {\bibfield  {journal}
  {\bibinfo  {journal} {Phys. Rev. Lett.}\ }\textbf {\bibinfo {volume} {121}},\
  \bibinfo {pages} {257001} (\bibinfo {year} {2018})}\BibitemShut {NoStop}%
\bibitem [{\citenamefont {Stauber}\ \emph {et~al.}(2018)\citenamefont
  {Stauber}, \citenamefont {Low},\ and\ \citenamefont
  {G\'omez-Santos}}]{2018Low}%
  \BibitemOpen
  \bibfield  {author} {\bibinfo {author} {\bibfnamefont {T.}~\bibnamefont
  {Stauber}}, \bibinfo {author} {\bibfnamefont {T.}~\bibnamefont {Low}},\ and\
  \bibinfo {author} {\bibfnamefont {G.}~\bibnamefont {G\'omez-Santos}},\ }\href
  {https://doi.org/10.1103/PhysRevLett.120.046801} {\bibfield  {journal}
  {\bibinfo  {journal} {Phys. Rev. Lett.}\ }\textbf {\bibinfo {volume} {120}},\
  \bibinfo {pages} {046801} (\bibinfo {year} {2018})}\BibitemShut {NoStop}%
\bibitem [{\citenamefont {{Pantale{\'o}n}}\ \emph {et~al.}(2022)\citenamefont
  {{Pantale{\'o}n}}, \citenamefont {{Tien Phong}}, \citenamefont {{Naumis}},\
  and\ \citenamefont {{Guinea}}}]{2022Pantaleon}%
  \BibitemOpen
  \bibfield  {author} {\bibinfo {author} {\bibfnamefont {P.~A.}\ \bibnamefont
  {{Pantale{\'o}n}}}, \bibinfo {author} {\bibfnamefont {V.}~\bibnamefont {{Tien
  Phong}}}, \bibinfo {author} {\bibfnamefont {G.~G.}\ \bibnamefont
  {{Naumis}}},\ and\ \bibinfo {author} {\bibfnamefont {F.}~\bibnamefont
  {{Guinea}}},\ }\href@noop {} {\bibfield  {journal} {\bibinfo  {journal}
  {arXiv e-prints}\ ,\ \bibinfo {eid} {arXiv:2204.09619}} (\bibinfo {year}
  {2022})},\ \Eprint {https://arxiv.org/abs/2204.09619} {arXiv:2204.09619
  [cond-mat.mes-hall]} \BibitemShut {NoStop}%
\bibitem [{\citenamefont {{Fu}}\ \emph {et~al.}(2020)\citenamefont {{Fu}},
  \citenamefont {{K{\"o}nig}}, \citenamefont {{Wilson}}, \citenamefont
  {{Chou}},\ and\ \citenamefont {{Pixley}}}]{2020Yixing}%
  \BibitemOpen
  \bibfield  {author} {\bibinfo {author} {\bibfnamefont {Y.}~\bibnamefont
  {{Fu}}}, \bibinfo {author} {\bibfnamefont {E.~J.}\ \bibnamefont
  {{K{\"o}nig}}}, \bibinfo {author} {\bibfnamefont {J.~H.}\ \bibnamefont
  {{Wilson}}}, \bibinfo {author} {\bibfnamefont {Y.-Z.}\ \bibnamefont
  {{Chou}}},\ and\ \bibinfo {author} {\bibfnamefont {J.~H.}\ \bibnamefont
  {{Pixley}}},\ }\href {https://doi.org/10.1038/s41535-020-00271-9} {\bibfield
  {journal} {\bibinfo  {journal} {npj Quantum Materials}\ }\textbf {\bibinfo
  {volume} {5}},\ \bibinfo {eid} {71} (\bibinfo {year} {2020})},\ \Eprint
  {https://arxiv.org/abs/1809.04604} {arXiv:1809.04604 [cond-mat.str-el]}
  \BibitemShut {NoStop}%
\bibitem [{\citenamefont {Bistritzer}\ and\ \citenamefont
  {MacDonald}(2011)}]{MacDonald2011}%
  \BibitemOpen
  \bibfield  {author} {\bibinfo {author} {\bibfnamefont {R.}~\bibnamefont
  {Bistritzer}}\ and\ \bibinfo {author} {\bibfnamefont {A.~H.}\ \bibnamefont
  {MacDonald}},\ }\href {https://doi.org/10.1073/pnas.1108174108} {\bibfield
  {journal} {\bibinfo  {journal} {Proceedings of the National Academy of
  Sciences}\ }\textbf {\bibinfo {volume} {108}},\ \bibinfo {pages} {12233}
  (\bibinfo {year} {2011})}\BibitemShut {NoStop}%
\bibitem [{\citenamefont {San-Jose}\ \emph {et~al.}(2012)\citenamefont
  {San-Jose}, \citenamefont {González},\ and\ \citenamefont
  {Guinea}}]{Guinea2012}%
  \BibitemOpen
  \bibfield  {author} {\bibinfo {author} {\bibfnamefont {P.}~\bibnamefont
  {San-Jose}}, \bibinfo {author} {\bibfnamefont {J.}~\bibnamefont
  {González}},\ and\ \bibinfo {author} {\bibfnamefont {F.}~\bibnamefont
  {Guinea}},\ }\href
  {https://doi.org/https://doi.org/10.1103/PhysRevLett.108.216802} {\bibfield
  {journal} {\bibinfo  {journal} {Phys. Rev. Lett.}\ }\textbf {\bibinfo
  {volume} {108}},\ \bibinfo {pages} {216802} (\bibinfo {year}
  {2012})}\BibitemShut {NoStop}%
\bibitem [{\citenamefont {Gerardo}\ \emph {et~al.}(2021)\citenamefont
  {Gerardo}, \citenamefont {Naumis}, \citenamefont {A.}, \citenamefont
  {Navarro-Labastida}, \citenamefont {Enrique}, \citenamefont
  {Aguilar-Méndez}, \citenamefont {Abdiel},\ and\ \citenamefont
  {Espinosa-Champo}}]{Naumis2021}%
  \BibitemOpen
  \bibfield  {author} {\bibinfo {author} {\bibnamefont {Gerardo}}, \bibinfo
  {author} {\bibfnamefont {G.}~\bibnamefont {Naumis}}, \bibinfo {author}
  {\bibfnamefont {L.}~\bibnamefont {A.}}, \bibinfo {author} {\bibnamefont
  {Navarro-Labastida}}, \bibinfo {author} {\bibnamefont {Enrique}}, \bibinfo
  {author} {\bibnamefont {Aguilar-Méndez}}, \bibinfo {author} {\bibnamefont
  {Abdiel}},\ and\ \bibinfo {author} {\bibnamefont {Espinosa-Champo}},\ }\href
  {https://doi.org/https://doi.org/10.1103/PhysRevB.103.245418} {\bibfield
  {journal} {\bibinfo  {journal} {Phys. Rev. B}\ }\textbf {\bibinfo {volume}
  {103}},\ \bibinfo {pages} {245418} (\bibinfo {year} {2021})}\BibitemShut
  {NoStop}%
\bibitem [{\citenamefont {Tarnopolsky}\ \emph {et~al.}(2019)\citenamefont
  {Tarnopolsky}, \citenamefont {Kruchkov},\ and\ \citenamefont
  {Vishwanath}}]{Tarnpolsky2019}%
  \BibitemOpen
  \bibfield  {author} {\bibinfo {author} {\bibfnamefont {G.}~\bibnamefont
  {Tarnopolsky}}, \bibinfo {author} {\bibfnamefont {A.~J.}\ \bibnamefont
  {Kruchkov}},\ and\ \bibinfo {author} {\bibfnamefont {A.}~\bibnamefont
  {Vishwanath}},\ }\href {https://doi.org/10.1103/PhysRevLett.122.106405}
  {\bibfield  {journal} {\bibinfo  {journal} {Phys. Rev. Lett.}\ }\textbf
  {\bibinfo {volume} {122}},\ \bibinfo {pages} {106405} (\bibinfo {year}
  {2019})}\BibitemShut {NoStop}%
\bibitem [{\citenamefont {Patrick}\ \emph {et~al.}(2020)\citenamefont
  {Patrick}, \citenamefont {J.}, \citenamefont {Ledwith}, \citenamefont
  {Grigory}, \citenamefont {Tarnopolsky}, \citenamefont {Eslam}, \citenamefont
  {Khalaf}, \citenamefont {Ashvin},\ and\ \citenamefont
  {Vishwanath}}]{Ledwidth2020}%
  \BibitemOpen
  \bibfield  {author} {\bibinfo {author} {\bibnamefont {Patrick}}, \bibinfo
  {author} {\bibnamefont {J.}}, \bibinfo {author} {\bibnamefont {Ledwith}},
  \bibinfo {author} {\bibnamefont {Grigory}}, \bibinfo {author} {\bibnamefont
  {Tarnopolsky}}, \bibinfo {author} {\bibnamefont {Eslam}}, \bibinfo {author}
  {\bibnamefont {Khalaf}}, \bibinfo {author} {\bibnamefont {Ashvin}},\ and\
  \bibinfo {author} {\bibnamefont {Vishwanath}},\ }\href
  {https://doi.org/https://doi.org/10.1103/PhysRevResearch.2.023237} {\bibfield
   {journal} {\bibinfo  {journal} {Phys. Rev. Research}\ }\textbf {\bibinfo
  {volume} {2}},\ \bibinfo {pages} {023237} (\bibinfo {year}
  {2020})}\BibitemShut {NoStop}%
\bibitem [{\citenamefont {Ledwith}\ \emph {et~al.}(2021)\citenamefont
  {Ledwith}, \citenamefont {Khalaf},\ and\ \citenamefont
  {Vishwanath}}]{Ledwith2021}%
  \BibitemOpen
  \bibfield  {author} {\bibinfo {author} {\bibfnamefont {P.~J.}\ \bibnamefont
  {Ledwith}}, \bibinfo {author} {\bibfnamefont {E.}~\bibnamefont {Khalaf}},\
  and\ \bibinfo {author} {\bibfnamefont {A.}~\bibnamefont {Vishwanath}},\
  }\href {https://doi.org/https://doi.org/10.1016/j.aop.2021.168646} {\bibfield
   {journal} {\bibinfo  {journal} {Annals of Physics}\ ,\ \bibinfo {pages}
  {168646}} (\bibinfo {year} {2021})}\BibitemShut {NoStop}%
\bibitem [{\citenamefont {Jie}\ \emph {et~al.}(2021)\citenamefont {Jie},
  \citenamefont {Wang}, \citenamefont {Yunqin}, \citenamefont {Zheng},
  \citenamefont {Andrew}, \citenamefont {J.}, \citenamefont {Millis},\ and\
  \citenamefont {Cano}}]{WANGG2021}%
  \BibitemOpen
  \bibfield  {author} {\bibinfo {author} {\bibnamefont {Jie}}, \bibinfo
  {author} {\bibnamefont {Wang}}, \bibinfo {author} {\bibnamefont {Yunqin}},
  \bibinfo {author} {\bibnamefont {Zheng}}, \bibinfo {author} {\bibnamefont
  {Andrew}}, \bibinfo {author} {\bibnamefont {J.}}, \bibinfo {author}
  {\bibnamefont {Millis}},\ and\ \bibinfo {author} {\bibfnamefont
  {J.}~\bibnamefont {Cano}},\ }\href
  {https://doi.org/https://doi.org/10.1103/PhysRevResearch.3.023155} {\bibfield
   {journal} {\bibinfo  {journal} {Phys. Rev. Research}\ }\textbf {\bibinfo
  {volume} {3}},\ \bibinfo {pages} {023155} (\bibinfo {year}
  {2021})}\BibitemShut {NoStop}%
\bibitem [{\citenamefont {Wang}\ \emph {et~al.}(2021)\citenamefont {Wang},
  \citenamefont {Cano}, \citenamefont {Millis}, \citenamefont {Liu},\ and\
  \citenamefont {Yang}}]{CANO2021}%
  \BibitemOpen
  \bibfield  {author} {\bibinfo {author} {\bibfnamefont {J.}~\bibnamefont
  {Wang}}, \bibinfo {author} {\bibfnamefont {J.}~\bibnamefont {Cano}}, \bibinfo
  {author} {\bibfnamefont {A.~J.}\ \bibnamefont {Millis}}, \bibinfo {author}
  {\bibfnamefont {Z.}~\bibnamefont {Liu}},\ and\ \bibinfo {author}
  {\bibfnamefont {B.}~\bibnamefont {Yang}},\ }\href
  {https://doi.org/10.1103/PhysRevLett.127.246403} {\bibfield  {journal}
  {\bibinfo  {journal} {Phys. Rev. Lett.}\ }\textbf {\bibinfo {volume} {127}},\
  \bibinfo {pages} {246403} (\bibinfo {year} {2021})}\BibitemShut {NoStop}%
\bibitem [{\citenamefont {Popov}\ and\ \citenamefont
  {Milekhin}(2021)}]{popovF2021}%
  \BibitemOpen
  \bibfield  {author} {\bibinfo {author} {\bibfnamefont {F.~K.}\ \bibnamefont
  {Popov}}\ and\ \bibinfo {author} {\bibfnamefont {A.}~\bibnamefont
  {Milekhin}},\ }\href {https://doi.org/10.1103/PhysRevB.103.155150} {\bibfield
   {journal} {\bibinfo  {journal} {Phys. Rev. B}\ }\textbf {\bibinfo {volume}
  {103}},\ \bibinfo {pages} {155150} (\bibinfo {year} {2021})}\BibitemShut
  {NoStop}%
\bibitem [{\citenamefont {Liu}\ \emph {et~al.}(2018)\citenamefont {Liu},
  \citenamefont {Zhang}, \citenamefont {Chen},\ and\ \citenamefont
  {Yang}}]{2018ChengCheng}%
  \BibitemOpen
  \bibfield  {author} {\bibinfo {author} {\bibfnamefont {C.-C.}\ \bibnamefont
  {Liu}}, \bibinfo {author} {\bibfnamefont {L.-D.}\ \bibnamefont {Zhang}},
  \bibinfo {author} {\bibfnamefont {W.-Q.}\ \bibnamefont {Chen}},\ and\
  \bibinfo {author} {\bibfnamefont {F.}~\bibnamefont {Yang}},\ }\href
  {https://doi.org/10.1103/PhysRevLett.121.217001} {\bibfield  {journal}
  {\bibinfo  {journal} {Phys. Rev. Lett.}\ }\textbf {\bibinfo {volume} {121}},\
  \bibinfo {pages} {217001} (\bibinfo {year} {2018})}\BibitemShut {NoStop}%
\bibitem [{\citenamefont {Onari}\ and\ \citenamefont
  {Kontani}(2022)}]{2022Onari}%
  \BibitemOpen
  \bibfield  {author} {\bibinfo {author} {\bibfnamefont {S.}~\bibnamefont
  {Onari}}\ and\ \bibinfo {author} {\bibfnamefont {H.}~\bibnamefont
  {Kontani}},\ }\href {https://doi.org/10.1103/PhysRevLett.128.066401}
  {\bibfield  {journal} {\bibinfo  {journal} {Phys. Rev. Lett.}\ }\textbf
  {\bibinfo {volume} {128}},\ \bibinfo {pages} {066401} (\bibinfo {year}
  {2022})}\BibitemShut {NoStop}%
\bibitem [{\citenamefont {Herrera}\ and\ \citenamefont
  {Naumis}(2021)}]{Saul2021}%
  \BibitemOpen
  \bibfield  {author} {\bibinfo {author} {\bibfnamefont {S.~A.}\ \bibnamefont
  {Herrera}}\ and\ \bibinfo {author} {\bibfnamefont {G.~G.}\ \bibnamefont
  {Naumis}},\ }\href {https://doi.org/10.1103/PhysRevB.104.115424} {\bibfield
  {journal} {\bibinfo  {journal} {Phys. Rev. B}\ }\textbf {\bibinfo {volume}
  {104}},\ \bibinfo {pages} {115424} (\bibinfo {year} {2021})}\BibitemShut
  {NoStop}%
\bibitem [{\citenamefont {De~Beule}\ \emph {et~al.}(2021)\citenamefont
  {De~Beule}, \citenamefont {Dominguez},\ and\ \citenamefont
  {Recher}}]{Patrick2021}%
  \BibitemOpen
  \bibfield  {author} {\bibinfo {author} {\bibfnamefont {C.}~\bibnamefont
  {De~Beule}}, \bibinfo {author} {\bibfnamefont {F.}~\bibnamefont
  {Dominguez}},\ and\ \bibinfo {author} {\bibfnamefont {P.}~\bibnamefont
  {Recher}},\ }\href {https://doi.org/10.1103/PhysRevB.104.195410} {\bibfield
  {journal} {\bibinfo  {journal} {Phys. Rev. B}\ }\textbf {\bibinfo {volume}
  {104}},\ \bibinfo {pages} {195410} (\bibinfo {year} {2021})}\BibitemShut
  {NoStop}%
\bibitem [{\citenamefont {Sheffer}\ and\ \citenamefont
  {Stern}(2021)}]{2021yardenn}%
  \BibitemOpen
  \bibfield  {author} {\bibinfo {author} {\bibfnamefont {Y.}~\bibnamefont
  {Sheffer}}\ and\ \bibinfo {author} {\bibfnamefont {A.}~\bibnamefont
  {Stern}},\ }\href {https://doi.org/10.1103/PhysRevB.104.L121405} {\bibfield
  {journal} {\bibinfo  {journal} {Phys. Rev. B}\ }\textbf {\bibinfo {volume}
  {104}},\ \bibinfo {pages} {L121405} (\bibinfo {year} {2021})}\BibitemShut
  {NoStop}%
\bibitem [{\citenamefont {Navarro-Labastida}\ \emph {et~al.}(2022)\citenamefont
  {Navarro-Labastida}, \citenamefont {Espinosa-Champo}, \citenamefont
  {Aguilar-Mendez},\ and\ \citenamefont {Naumis}}]{Naumis2022}%
  \BibitemOpen
  \bibfield  {author} {\bibinfo {author} {\bibfnamefont {L.~A.}\ \bibnamefont
  {Navarro-Labastida}}, \bibinfo {author} {\bibfnamefont {A.}~\bibnamefont
  {Espinosa-Champo}}, \bibinfo {author} {\bibfnamefont {E.}~\bibnamefont
  {Aguilar-Mendez}},\ and\ \bibinfo {author} {\bibfnamefont {G.~G.}\
  \bibnamefont {Naumis}},\ }\href {https://doi.org/10.1103/PhysRevB.105.115434}
  {\bibfield  {journal} {\bibinfo  {journal} {Phys. Rev. B}\ }\textbf {\bibinfo
  {volume} {105}},\ \bibinfo {pages} {115434} (\bibinfo {year}
  {2022})}\BibitemShut {NoStop}%
\bibitem [{\citenamefont {{Matsumoto}}\ \emph {et~al.}(2022)\citenamefont
  {{Matsumoto}}, \citenamefont {{Mizoguchi}},\ and\ \citenamefont
  {{Hatsugai}}}]{2022Mizoguchi}%
  \BibitemOpen
  \bibfield  {author} {\bibinfo {author} {\bibfnamefont {D.}~\bibnamefont
  {{Matsumoto}}}, \bibinfo {author} {\bibfnamefont {T.}~\bibnamefont
  {{Mizoguchi}}},\ and\ \bibinfo {author} {\bibfnamefont {Y.}~\bibnamefont
  {{Hatsugai}}},\ }\href@noop {} {\bibfield  {journal} {\bibinfo  {journal}
  {arXiv e-prints}\ ,\ \bibinfo {eid} {arXiv:2207.14540}} (\bibinfo {year}
  {2022})},\ \Eprint {https://arxiv.org/abs/2207.14540} {arXiv:2207.14540
  [cond-mat.mes-hall]} \BibitemShut {NoStop}%
\bibitem [{\citenamefont {Mizoguchi}\ \emph {et~al.}(2021)\citenamefont
  {Mizoguchi}, \citenamefont {Yoshida},\ and\ \citenamefont
  {Hatsugai}}]{Hatsugai2021}%
  \BibitemOpen
  \bibfield  {author} {\bibinfo {author} {\bibfnamefont {T.}~\bibnamefont
  {Mizoguchi}}, \bibinfo {author} {\bibfnamefont {T.}~\bibnamefont {Yoshida}},\
  and\ \bibinfo {author} {\bibfnamefont {Y.}~\bibnamefont {Hatsugai}},\ }\href
  {https://doi.org/10.1103/PhysRevB.103.045136} {\bibfield  {journal} {\bibinfo
   {journal} {Phys. Rev. B}\ }\textbf {\bibinfo {volume} {103}},\ \bibinfo
  {pages} {045136} (\bibinfo {year} {2021})}\BibitemShut {NoStop}%
\bibitem [{\citenamefont {{Roychowdhury}}\ \emph {et~al.}(2022)\citenamefont
  {{Roychowdhury}}, \citenamefont {{Attig}}, \citenamefont {{Trebst}},\ and\
  \citenamefont {{Lawler}}}]{2022Krishanu}%
  \BibitemOpen
  \bibfield  {author} {\bibinfo {author} {\bibfnamefont {K.}~\bibnamefont
  {{Roychowdhury}}}, \bibinfo {author} {\bibfnamefont {J.}~\bibnamefont
  {{Attig}}}, \bibinfo {author} {\bibfnamefont {S.}~\bibnamefont {{Trebst}}},\
  and\ \bibinfo {author} {\bibfnamefont {M.~J.}\ \bibnamefont {{Lawler}}},\
  }\href@noop {} {\bibfield  {journal} {\bibinfo  {journal} {arXiv e-prints}\
  ,\ \bibinfo {eid} {arXiv:2207.09475}} (\bibinfo {year} {2022})},\ \Eprint
  {https://arxiv.org/abs/2207.09475} {arXiv:2207.09475 [cond-mat.str-el]}
  \BibitemShut {NoStop}%
\bibitem [{\citenamefont {Eslam}\ \emph {et~al.}(2019)\citenamefont {Eslam},
  \citenamefont {Khalaf}, \citenamefont {J.}, \citenamefont {Kruchkov},
  \citenamefont {Grigory}, \citenamefont {Tarnopolsky}, \citenamefont
  {Ashvin},\ and\ \citenamefont {Vishwanath}}]{Khalaff2019}%
  \BibitemOpen
  \bibfield  {author} {\bibinfo {author} {\bibnamefont {Eslam}}, \bibinfo
  {author} {\bibnamefont {Khalaf}}, \bibinfo {author} {\bibfnamefont
  {A.}~\bibnamefont {J.}}, \bibinfo {author} {\bibnamefont {Kruchkov}},
  \bibinfo {author} {\bibnamefont {Grigory}}, \bibinfo {author} {\bibnamefont
  {Tarnopolsky}}, \bibinfo {author} {\bibnamefont {Ashvin}},\ and\ \bibinfo
  {author} {\bibnamefont {Vishwanath}},\ }\href
  {https://doi.org/https://doi.org/10.1103/PhysRevB.100.085109} {\bibfield
  {journal} {\bibinfo  {journal} {Phys. Rev. B}\ }\textbf {\bibinfo {volume}
  {100}},\ \bibinfo {pages} {085109} (\bibinfo {year} {2019})}\BibitemShut
  {NoStop}%
\bibitem [{\citenamefont {Hofstadter}(1976)}]{Hofstadter_1976}%
  \BibitemOpen
  \bibfield  {author} {\bibinfo {author} {\bibfnamefont {D.~R.}\ \bibnamefont
  {Hofstadter}},\ }\href {https://doi.org/10.1103/PhysRevB.14.2239} {\bibfield
  {journal} {\bibinfo  {journal} {Phys. Rev. B}\ }\textbf {\bibinfo {volume}
  {14}},\ \bibinfo {pages} {2239} (\bibinfo {year} {1976})}\BibitemShut
  {NoStop}%
\bibitem [{\citenamefont {{Wang}}\ \emph
  {et~al.}(2020{\natexlab{b}})\citenamefont {{Wang}}, \citenamefont {{Zheng}},
  \citenamefont {{Millis}},\ and\ \citenamefont {{Cano}}}]{Wang2020}%
  \BibitemOpen
  \bibfield  {author} {\bibinfo {author} {\bibfnamefont {J.}~\bibnamefont
  {{Wang}}}, \bibinfo {author} {\bibfnamefont {Y.}~\bibnamefont {{Zheng}}},
  \bibinfo {author} {\bibfnamefont {A.~J.}\ \bibnamefont {{Millis}}},\ and\
  \bibinfo {author} {\bibfnamefont {J.}~\bibnamefont {{Cano}}},\ }\href@noop {}
  {\bibfield  {journal} {\bibinfo  {journal} {arXiv e-prints}\ ,\ \bibinfo
  {eid} {arXiv:2010.03589}} (\bibinfo {year} {2020}{\natexlab{b}})},\ \Eprint
  {https://arxiv.org/abs/2010.03589} {arXiv:2010.03589 [cond-mat.mes-hall]}
  \BibitemShut {NoStop}%
\bibitem [{\citenamefont {Roscoe~B.}(2010)}]{BookDE}%
  \BibitemOpen
  \bibfield  {author} {\bibinfo {author} {\bibfnamefont {W.}~\bibnamefont
  {Roscoe~B.}},\ }\href@noop {} {\emph {\bibinfo {title} {Asymptotic Analysis
  of Differential Equations}}}\ (\bibinfo  {publisher} {Imperial College
  Press},\ \bibinfo {year} {2010})\BibitemShut {NoStop}%
\bibitem [{\citenamefont {Greiner}(2001)}]{BookQM}%
  \BibitemOpen
  \bibfield  {author} {\bibinfo {author} {\bibfnamefont {W.}~\bibnamefont
  {Greiner}},\ }\href@noop {} {\emph {\bibinfo {title} {Quantum Mechanics}}}\
  (\bibinfo  {publisher} {Springer},\ \bibinfo {year} {2001})\BibitemShut
  {NoStop}%
\bibitem [{\citenamefont {Girvin}\ and\ \citenamefont
  {Yang}(2019)}]{BookGrivin}%
  \BibitemOpen
  \bibfield  {author} {\bibinfo {author} {\bibfnamefont {S.~M.}\ \bibnamefont
  {Girvin}}\ and\ \bibinfo {author} {\bibfnamefont {K.}~\bibnamefont {Yang}},\
  }\href@noop {} {\emph {\bibinfo {title} {Modern Condensed Matter Physics}}}\
  (\bibinfo  {publisher} {Cambridge University Press},\ \bibinfo {year}
  {2019})\BibitemShut {NoStop}%
\bibitem [{\citenamefont {Cohen-Tannoudji}(1991)}]{BookCohen}%
  \BibitemOpen
  \bibfield  {author} {\bibinfo {author} {\bibfnamefont {C.}~\bibnamefont
  {Cohen-Tannoudji}},\ }\href@noop {} {\emph {\bibinfo {title} {Quantum
  Mechanics}}}\ (\bibinfo  {publisher} {Wiley},\ \bibinfo {year}
  {1991})\BibitemShut {NoStop}%
\bibitem [{\citenamefont {Kirkpatrick}\ and\ \citenamefont
  {Eggarter}(1972)}]{Kirkpatrick1972}%
  \BibitemOpen
  \bibfield  {author} {\bibinfo {author} {\bibfnamefont {S.}~\bibnamefont
  {Kirkpatrick}}\ and\ \bibinfo {author} {\bibfnamefont {T.~P.}\ \bibnamefont
  {Eggarter}},\ }\href {https://doi.org/10.1103/PhysRevB.6.3598} {\bibfield
  {journal} {\bibinfo  {journal} {Phys. Rev. B}\ }\textbf {\bibinfo {volume}
  {6}},\ \bibinfo {pages} {3598} (\bibinfo {year} {1972})}\BibitemShut
  {NoStop}%
\bibitem [{\citenamefont {Naumis}\ \emph {et~al.}(2002)\citenamefont {Naumis},
  \citenamefont {Wang},\ and\ \citenamefont {Barrio}}]{Naumis2002}%
  \BibitemOpen
  \bibfield  {author} {\bibinfo {author} {\bibfnamefont {G.~G.}\ \bibnamefont
  {Naumis}}, \bibinfo {author} {\bibfnamefont {C.}~\bibnamefont {Wang}},\ and\
  \bibinfo {author} {\bibfnamefont {R.~A.}\ \bibnamefont {Barrio}},\ }\href
  {https://doi.org/10.1103/PhysRevB.65.134203} {\bibfield  {journal} {\bibinfo
  {journal} {Phys. Rev. B}\ }\textbf {\bibinfo {volume} {65}},\ \bibinfo
  {pages} {134203} (\bibinfo {year} {2002})}\BibitemShut {NoStop}%
\bibitem [{\citenamefont {Barrios-Vargas}\ and\ \citenamefont
  {Naumis}(2013)}]{Barrios2013}%
  \BibitemOpen
  \bibfield  {author} {\bibinfo {author} {\bibfnamefont {J.}~\bibnamefont
  {Barrios-Vargas}}\ and\ \bibinfo {author} {\bibfnamefont {G.~G.}\
  \bibnamefont {Naumis}},\ }\href
  {https://doi.org/https://doi.org/10.1016/j.ssc.2013.03.006} {\bibfield
  {journal} {\bibinfo  {journal} {Solid State Communications}\ }\textbf
  {\bibinfo {volume} {162}},\ \bibinfo {pages} {23} (\bibinfo {year}
  {2013})}\BibitemShut {NoStop}%
\bibitem [{\citenamefont {Huerta}\ and\ \citenamefont
  {Naumis}(2002)}]{Huerta2002PLA}%
  \BibitemOpen
  \bibfield  {author} {\bibinfo {author} {\bibfnamefont {A.}~\bibnamefont
  {Huerta}}\ and\ \bibinfo {author} {\bibfnamefont {G.}~\bibnamefont
  {Naumis}},\ }\href
  {https://doi.org/https://doi.org/10.1016/S0375-9601(02)00519-4} {\bibfield
  {journal} {\bibinfo  {journal} {Physics Letters A}\ }\textbf {\bibinfo
  {volume} {299}},\ \bibinfo {pages} {660} (\bibinfo {year}
  {2002})}\BibitemShut {NoStop}%
\bibitem [{\citenamefont {Huerta}\ \emph {et~al.}(2004)\citenamefont {Huerta},
  \citenamefont {Naumis}, \citenamefont {Wasan}, \citenamefont {Henderson},\
  and\ \citenamefont {Trokhymchuk}}]{Huerta2004}%
  \BibitemOpen
  \bibfield  {author} {\bibinfo {author} {\bibfnamefont {A.}~\bibnamefont
  {Huerta}}, \bibinfo {author} {\bibfnamefont {G.~G.}\ \bibnamefont {Naumis}},
  \bibinfo {author} {\bibfnamefont {D.~T.}\ \bibnamefont {Wasan}}, \bibinfo
  {author} {\bibfnamefont {D.}~\bibnamefont {Henderson}},\ and\ \bibinfo
  {author} {\bibfnamefont {A.}~\bibnamefont {Trokhymchuk}},\ }\href
  {https://doi.org/10.1063/1.1632893} {\bibfield  {journal} {\bibinfo
  {journal} {The Journal of Chemical Physics}\ }\textbf {\bibinfo {volume}
  {120}},\ \bibinfo {pages} {1506} (\bibinfo {year} {2004})},\ \Eprint
  {https://arxiv.org/abs/https://doi.org/10.1063/1.1632893}
  {https://doi.org/10.1063/1.1632893} \BibitemShut {NoStop}%
\bibitem [{\citenamefont {Flores-Ruiz}\ \emph {et~al.}(2010)\citenamefont
  {Flores-Ruiz}, \citenamefont {Naumis},\ and\ \citenamefont
  {Phillips}}]{Flores2010}%
  \BibitemOpen
  \bibfield  {author} {\bibinfo {author} {\bibfnamefont {H.~M.}\ \bibnamefont
  {Flores-Ruiz}}, \bibinfo {author} {\bibfnamefont {G.~G.}\ \bibnamefont
  {Naumis}},\ and\ \bibinfo {author} {\bibfnamefont {J.~C.}\ \bibnamefont
  {Phillips}},\ }\href {https://doi.org/10.1103/PhysRevB.82.214201} {\bibfield
  {journal} {\bibinfo  {journal} {Phys. Rev. B}\ }\textbf {\bibinfo {volume}
  {82}},\ \bibinfo {pages} {214201} (\bibinfo {year} {2010})}\BibitemShut
  {NoStop}%
\bibitem [{\citenamefont {{Navarro-Labastida}}\ \emph
  {et~al.}(2021)\citenamefont {{Navarro-Labastida}}, \citenamefont
  {{Dom{\'\i}nguez-Serna}},\ and\ \citenamefont {{Rojas}}}]{NavarroL2021}%
  \BibitemOpen
  \bibfield  {author} {\bibinfo {author} {\bibfnamefont {L.~A.}\ \bibnamefont
  {{Navarro-Labastida}}}, \bibinfo {author} {\bibfnamefont {F.~A.}\
  \bibnamefont {{Dom{\'\i}nguez-Serna}}},\ and\ \bibinfo {author}
  {\bibfnamefont {F.}~\bibnamefont {{Rojas}}},\ }\href
  {https://doi.org/10.48550/arXiv.2106.02756} {\bibfield  {journal} {\bibinfo
  {journal} {arXiv e-prints}\ ,\ \bibinfo {eid} {arXiv:2106.02756}} (\bibinfo
  {year} {2021})},\ \Eprint {https://arxiv.org/abs/2106.02756}
  {arXiv:2106.02756 [quant-ph]} \BibitemShut {NoStop}%
\bibitem [{\citenamefont {Flores-Ruiz}\ and\ \citenamefont
  {Naumis}(2011)}]{Flores2011}%
  \BibitemOpen
  \bibfield  {author} {\bibinfo {author} {\bibfnamefont {H.~M.}\ \bibnamefont
  {Flores-Ruiz}}\ and\ \bibinfo {author} {\bibfnamefont {G.~G.}\ \bibnamefont
  {Naumis}},\ }\href {https://doi.org/10.1103/PhysRevB.83.184204} {\bibfield
  {journal} {\bibinfo  {journal} {Phys. Rev. B}\ }\textbf {\bibinfo {volume}
  {83}},\ \bibinfo {pages} {184204} (\bibinfo {year} {2011})}\BibitemShut
  {NoStop}%
\bibitem [{\citenamefont {Moukarzel}\ and\ \citenamefont
  {Naumis}(2022)}]{Moukarzel_2022}%
  \BibitemOpen
  \bibfield  {author} {\bibinfo {author} {\bibfnamefont {C.~F.}\ \bibnamefont
  {Moukarzel}}\ and\ \bibinfo {author} {\bibfnamefont {G.~G.}\ \bibnamefont
  {Naumis}},\ }\href {https://doi.org/10.1103/PhysRevE.106.035001} {\bibfield
  {journal} {\bibinfo  {journal} {Phys. Rev. E}\ }\textbf {\bibinfo {volume}
  {106}},\ \bibinfo {pages} {035001} (\bibinfo {year} {2022})}\BibitemShut
  {NoStop}%
\bibitem [{\citenamefont {Friis}\ \emph {et~al.}(2015)\citenamefont {Friis},
  \citenamefont {Melnikov}, \citenamefont {Kirchmair},\ and\ \citenamefont
  {Briegel}}]{Friis2015}%
  \BibitemOpen
  \bibfield  {author} {\bibinfo {author} {\bibfnamefont {N.}~\bibnamefont
  {Friis}}, \bibinfo {author} {\bibfnamefont {A.~A.}\ \bibnamefont {Melnikov}},
  \bibinfo {author} {\bibfnamefont {G.}~\bibnamefont {Kirchmair}},\ and\
  \bibinfo {author} {\bibfnamefont {H.~J.}\ \bibnamefont {Briegel}},\ }\href
  {https://doi.org/10.1038/srep18036} {\bibfield  {journal} {\bibinfo
  {journal} {Scientific Reports}\ }\textbf {\bibinfo {volume} {5}},\ \bibinfo
  {pages} {18036} (\bibinfo {year} {2015})}\BibitemShut {NoStop}%
\bibitem [{\citenamefont {Betancur-Ocampo}\ \emph {et~al.}(2019)\citenamefont
  {Betancur-Ocampo}, \citenamefont {Leyvraz},\ and\ \citenamefont
  {Stegmann}}]{Betancur-Ocampo2019}%
  \BibitemOpen
  \bibfield  {author} {\bibinfo {author} {\bibfnamefont {Y.}~\bibnamefont
  {Betancur-Ocampo}}, \bibinfo {author} {\bibfnamefont {F.}~\bibnamefont
  {Leyvraz}},\ and\ \bibinfo {author} {\bibfnamefont {T.}~\bibnamefont
  {Stegmann}},\ }\href {https://doi.org/10.1021/acs.nanolett.9b02720}
  {\bibfield  {journal} {\bibinfo  {journal} {Nano Letters}\ }\textbf {\bibinfo
  {volume} {19}},\ \bibinfo {pages} {7760} (\bibinfo {year}
  {2019})}\BibitemShut {NoStop}%
\end{thebibliography}

%

\end{document}